\documentclass[preprint,12pt,authoryear]{elsarticle}
\usepackage{graphicx} %
\usepackage{natbib}
\usepackage[utf8]{inputenc} %
\usepackage[T1]{fontenc}    %
\usepackage{lmodern}
\usepackage[hidelinks]{hyperref}       %
\usepackage{url}            %
\usepackage{booktabs}       %
\usepackage{amsfonts}       %
\usepackage{nicefrac}       %
\usepackage{microtype}      %
\usepackage{xcolor}         %

\usepackage{amsmath,amssymb,amsthm}
\usepackage[ruled,vlined]{algorithm2e}
\SetAlgoCaptionLayout{justified}

\DeclareMathOperator*{\argmin}{arg\,min}

\newcommand{\BigO}{\mathcal O}

\newcommand{\Regret}{\mathrm{Regret}}
\newcommand{\cost}{\mathrm{cost}}
\DeclareMathOperator{\nnz}{nnz}

\newcommand{\R}{\mathbb R}

\usepackage{scalerel,mathtools,color}
\let\svsqrt\sqrt
\newsavebox\Nsqrt
\def\sr#1{\ThisStyle{%
		\savebox\Nsqrt{\scalebox{.5}[1]{$\SavedStyle\svsqrt{\phantom{\cramped{#1#1}}}$}}%
		\ooalign{\usebox{\Nsqrt}\cr\kern.2pt\usebox{\Nsqrt}\cr\hfil$\SavedStyle\cramped{#1}$}}}

\usepackage{enumitem}
\usepackage{bm}
\def\*#1{\mathbf{#1}}

\usepackage{threeparttable}
\usepackage{multirow}

\usepackage{grffile}

\newcommand\arxiv

\newcommand{\revision}[1]{#1}

\begin{document}

\begin{frontmatter}
\title{One-shot acceleration of transient PDE solvers\\via online-learned preconditioners\looseness-1
}

\ifdefined\anon\else
\author[1]{Mikhail Khodak}
\affiliation[1]{organization={University of Wisconsin-Madison},country={United States}}
\ifdefined\anon\else\ead{khodak@wisc.edu}\fi

\author[2]{Min Ki Jung}
\affiliation[2]{organization={Seoul National University},country={South Korea}}

\author[3]{Brian Wynne}
\affiliation[3]{organization={Princeton University},country={United States}}

\author[4]{\\Edmond Chow}
\affiliation[4]{organization={Georgia Institute of Technology},country={United States}}

\author[3]{Egemen Kolemen}
\fi

\begin{abstract}
    
    Data-driven acceleration of scientific computing workflows has been a high-profile aim of machine learning~(ML) for science, with numerical simulation of transient partial differential equations~(PDEs) being one of the main applications.
    The focus thus far has been on methods that require classical simulations to train, which when combined with the data-hungriness and optimization challenges of neural networks has caused difficulties in demonstrating a convincing advantage against strong classical baselines.
    We consider an alternative paradigm in which the learner uses a classical solver's own data to accelerate it, enabling a one-shot speedup of the simulation.
    Concretely, since transient PDEs often require solving a sequence of related linear systems, the feedback from repeated calls to a linear solver such as preconditioned conjugate gradient~(PCG) can be used by a bandit algorithm to online-learn an adaptive sequence of solver configurations~(e.g. preconditioners).
    The method we develop, PCGBandit, is implemented directly on top of the popular open-source software OpenFOAM, which we use to show its effectiveness on a set of fluid and magnetohydrodynamics~(MHD) problems.\looseness-1
    
\end{abstract}

\ifdefined\arxiv\else
\fi

\end{frontmatter}

\newpage
\section{Introduction}

Simulating the evolution of transient PDEs is one of the main tasks in scientific computing, with applications in inverse problems~\citep{isakov2017}, experimental and engineering design~\citep{borggaard1997pde,huan2024optimal}, and forecasting~\citep{kalnay2002atmospheric}.
In many cases of interest the computational challenges involved often incur significant costs or preclude simulations of sufficient accuracy~\citep{verstappen1997dns,reynolds2006fully}.
As a result, there has been significant interest in applying recent advances in deep learning to accelerating or even replacing classical solvers~\citep{han2018solving}.
High-profile approaches include physics-informed neural networks~(PINNs), in which the solution is a neural network found by minimizing the residual~\citep{raissi2019physics}, and neural operators, in which an ML model is trained on data generated by classical solvers in order to simulate their behavior~\citep{li2021fno}.
However, efforts in these directions continue to face optimization, sample complexity, and correctness issues, with recent meta-analyses finding underwhelming improvements relative to classical solvers~\citep{grossmann2024physics,mcgreivy2024weak}.

\begin{figure}[!t]
	\includegraphics[width=\linewidth]{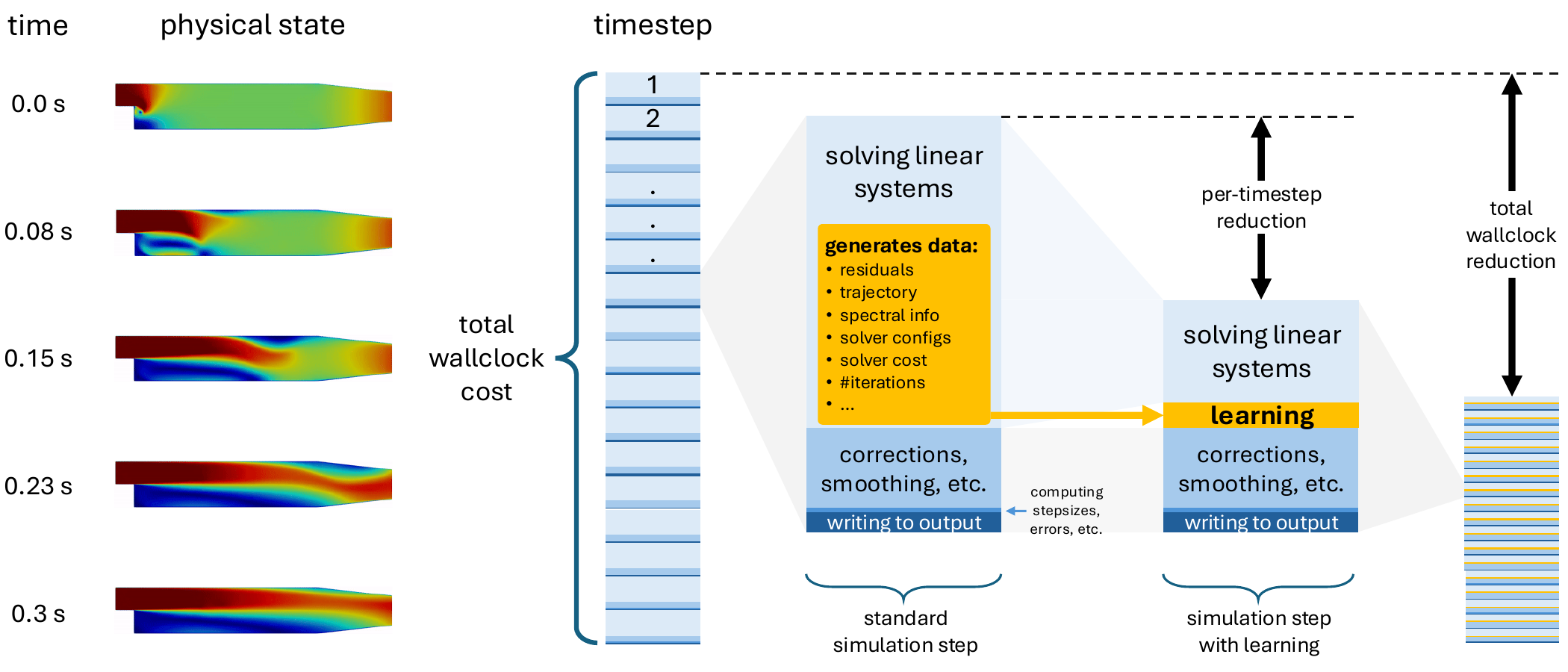}\ifdefined\arxiv\vspace{-4mm}\fi
	\caption{
		Illustration of a learning-enhanced numerical simulation setup.
		In transient simulations the wallclock cost is the sum of costs incurred at each in a sequence of discrete timesteps.
		For many practical cases, the cost at each step is dominated by solving one or more linear systems, with each solve generating data such as the solver settings used and the time needed to converge.
		By passing this feedback to a learning algorithm that configures the solver, we hope to reduce the cost of solving the linear systems by more than the cost of the extra computations performed by the learner.
		If so the total wallclock cost will be reduced in a one-shot manner, i.e. where learning occurs solely within the simulation itself.\looseness-1
	}
	\label{fig:overview}
\end{figure}

We consider an alternative paradigm of learning-enhanced numerical simulation, in which the optimization algorithms have guarantees, the number of samples required is zero, and correctness is inherited from classical solvers.
This approach, illustrated in Figure~\ref{fig:overview}, starts by identifying a repeated computation such as linear system solving that takes up a significant quantity of the cost;
it then uses sequential data generated by doing these computations to speed them up on future instances, e.g. by learning better preconditioners or components of preconditioners.\footnote{
	This is useful even when a {\em class} of solvers is known to be {\em asymptotically} optimal for the setting at hand~(e.g. multigrid for elliptic PDEs~\citep{hackbusch1985multigrid}), as the non-asymptotic cost may still depend heavily and unpredictably on the exact configuration within that class~(e.g. on the choice of smoother or interpolation operator in multigrid).
	For some parameters the optimal value may even depend on the computer system being used.
	\looseness-1} 
As shown by \citet{khodak2024learning}, standard adversarial bandit algorithms~\citep{auer2002exp3,zimmert2021tsallis,foster2020squarecb} can provably learn to configure certain linear solvers using only the number of iterations as feedback;
such learners are lightweight and can be directly extended to configure any solver and to use any type of feedback, including e.g. wallclock cost.
Since we run no additional simulations other than the one being accelerated, and because correctness is guaranteed so long as that of the learning-free simulation is,\footnote{\revision{This assumes simulation correctness is not affected by the choice of preconditioner.}} \revision{our} learning-enhanced methods can \revision{also} be easily compared directly to classical approaches.

To demonstrate these advantages, we develop a practical preconditioner-learning method and implement it  directly on top of the widely used numerical simulation software OpenFOAM~\citep{jasak2007openfoam}.
Our algorithm, PCGBandit,\ifdefined\anon\footnote{\url{www.anonymous.url}}\else\footnote{\revision{\url{https://github.com/mkhodak/PCGBandit}}}
\fi
works as a drop-in substitute for OpenFOAM's main symmetric positive-definite~(SPD) system solver PCG~(preconditioned conjugate gradient) that adaptively sets the preconditioner as the simulation progresses.
The underlying learning procedure is lightweight, with a computational overhead independent of the problem size.
We use it to learn multigrid and incomplete Cholesky preconditioners for four OpenFOAM tutorial simulations and two problems from the open-source MHD code FreeMHD~\citep{wynne2025freemhd}, which is built on top of OpenFOAM.
Across these evaluations, PCGBandit obtains up to 1.5$\times$ reductions in wallclock time relative to fixed baseline preconditioners while always being at least as fast as default settings.

\section{Related work}

Our contribution builds upon the theoretical study by \citet{khodak2024learning} of online-learned preconditioners.
On a technical level, we consider multiple different learning procedures from the bandit literature and settle on a different, variance-reduced estimation technique~\citep{zimmert2021tsallis} to approximate solver costs that leads to improved empirical performance.
Furthermore, we implement our approach in a widely used numerical software, demonstrating the paradigm's practical utility and broad applicability beyond toy settings.
Prior to that work, the underlying idea of using adversarial bandit learners to do online algorithm configuration was studied by \citet{gagliolo2006learning,gagliolo2011algorithm}, albeit mainly in the context of combinatorial optimization.\looseness-1

Within numerical simulation, \citet{pawar2021distributed} and \citet{geise2023learning} studied the use of online learning---specifically deep reinforcement learning~(RL)---for configuring transient PDE solvers, with the latter also focusing specifically on OpenFOAM.
Their approaches involved pretraining RL agents on thousands of simulations, which is expensive and potentially non-transferable to different simulation settings.
Similar pretraining of {\em static} preconditioners for conjugate gradient was also studied by \citet{li2023learning} and \citet{kopani2025deeponet}.
In contrast to these studies, we use a one-shot approach in which only the data from the simulation at hand is used to accelerate it.
\revision{Aside from \citet{khodak2024learning}, the most closely related work is that of \citet{mishra2024online}, who used online Bayesian optimization to tune polynomial multigrid cycles during a fluid simulation. Our approach differs in the numerical subroutine it targets and in its learning methodology, favoring a bandit approach that is explicitly robust to the non-stationary cost sequences found in numerical simulations, a property that does not typically hold for Bayesian methods.}
We are aware of one other effort at such {\em within-simulation} learning, by \citet{sirignano2023dynamic}, who use online gradient descent to learn a turbulence closure model.
This approach can reduce the number of mesh points required for accurate simulation, but its correctness is not guaranteed and costly to verify, requiring a comparison of the steady-state statistics to those of a fully resolved simulation.
In contrast, our approach accelerates learning-free simulation at the same resolution while inheriting its correctness guarantees across the entire transient trajectory.\looseness-1

Beyond learning, the acceleration of linear solvers across sequences of related instances has also seen significant study within the scientific computing community~\citep{anzt2016updating,austin2021initial,bellavia2011efficient,bergamaschi2020survey,bertaccini2004efficient,giraud2007incremental,giraud2022block,soodhalter2014krylov,tebbens2007efficient,elbouyahyaoui2021restarted,guido2024subspace}.
A method like PCGBandit is orthogonal to these approaches and makes limited assumptions on the underlying sequence of systems, so their combined use can potentially result in compounded improvements.

\begin{algorithm}[!t]
	\begin{minipage}{\linewidth}
		\caption{\label{alg:bandit}
		Comparison between the proposed adversarial bandit approach~(left) and a numerical simulation code that adaptively sets solver configurations~(right), with colors denoting correspondence between specific steps.
		The performance of a bandit method is measured via its regret relative to the best fixed action in hindsight, whereas a simulation code is evaluated by its wallclock.
		These metrics differ by an additive constant.\looseness-1
	}
	\end{minipage}
	\begin{minipage}{0.42\linewidth}
		\DontPrintSemicolon
		$d\gets$ number of actions\;
		\vspace{-11pt}\;
		{\color{teal} adversary chooses cost functions $\cost_1,\dots,\cost_T:[d]\mapsto\R$}\;
		\For{round $t=1,\dots,T$}{
			{\color{cyan} agent picks action $i_t\in[d]$}\;
			{\color{purple} adversary reveals $\cost_t(i_t)$}\;
		}
		\vspace{-10pt}\;
		{\bf Performance measure:}\\
		regret\\ $=\sum\limits_{t=1}^T\cost_t(i_t)-\min\limits_{i\in[d]}\sum\limits_{t=1}^T\cost_t(i)$
	\end{minipage}
	\hfill
	\begin{minipage}{0.42\linewidth}
		\DontPrintSemicolon
		$d\gets$ number of solver configs\;
		\vspace{-11pt}\;
		\For{timestep $t=1,\dots,T$}{
			{\color{teal} simulation requires solving a linear system $\*A_t\*x=\*b_t$}\;
			{\color{cyan} code picks solver $i_t\in[d]$}\;
			{\color{purple} code observes time $\cost_t(i_t)$ taken to solve $\*A_t\*x=\*b_t$}\;
		}
		\vspace{-9pt}\;
		{\bf Performance measure:}\vspace{4pt}\\
		wallclock $=\sum\limits_{t=1}^T\cost_t(i_t)$
	\end{minipage}
\end{algorithm}

\section{Methodology}

We consider numerical simulations where the main computational cost is in solving a sequence $\{(\*A_t,\*b_t)\}_{t=1}^T$ of $T$ linear systems $\*A_t\*x=\*b_t$, for SPD matrices $\*A_t\in\R^{n\times n}$ and target vectors $\*b_t\in\R^n$.
Here {\em solving} refers to finding a solution vector $\*x\in\R^n$ that satisfies a~(potentially $t$-dependent) convergence criterion, e.g. $\|\*b_t-\*A_t\*x\|_2\le\varepsilon\|\*b_t\|_2$ for some $\varepsilon>0$; for generality, we do not specify this formally, to account for more involved criteria.\footnote{\revision{\url{https://doc.cfd.direct/notes/cfd-general-principles/residual}}}
To find such solutions we will use conjugate gradient preconditioned with one of $d$ possible preconditioners, each of which specifies a function that constructs and applies an $\*A_t$-dependent SPD transformation of the residual at each iteration of PCG.
At each timestep $t\in[T]$ we select a preconditioner $i_t\in[d]$, solve the linear system $\*A_t\*x=\*b_t$, and receive feedback $\cost_t(i_t)\in\R_{\ge0}$ corresponding to the time required for preconditioner $i_t$ to solve linear system $t$.
The goal is to sequentially select preconditioners $\{i_t\}_{t=1}^T$ such that the total cost $\sum_{t=1}^T\cost_t(i_t)$ incurred is small.

\subsection{Solver configuration as an adversarial bandit problem}

As first observed in \citet{khodak2024learning} and formalized via the pseudo-code in Algorithm~\ref{alg:bandit}, our motivating observation is that the setup described above maps closely to {\bf online learning}~\citep{cesa-bianchi2006prediction}, a well-studied subfield of machine learning in which an agent (in our case, the simulation code) makes decisions (picks solvers) across a sequence of $T$ rounds (timesteps), with the goal of minimizing the total cost~(wallclock) of those decisions.
Since doing the latter is generally hard, the field has adopted a relative metric called {\bf regret}, in which the agent's performance is measured against that of the best {\em fixed} action in hindsight:
\begin{equation}\label{eq:regret}
	\Regret_T
	=\sum_{t=1}^T\cost_t(i_t)-\min_{i\in[d]}\sum_{t=1}^T\cost_t(i)
\end{equation}
An agent is said to successfully {\em learn} in this setting if the regret is {\em sublinear} in $T$, because in that case as $T\to\infty$ the average cost incurred per round approaches that of the best fixed action in hindsight.
In the case where the costs come from a fixed sequence of linear systems $\*A_t\*x=\*b_t$, successful learning implies that the average cost we incur is {\em asymptotically} (as $T\to\infty$) equal to that of the best solver configuration:
\begin{equation}\label{eq:wallclock}
	\underbrace{\frac1T\sum_{t=1}^T\cost_t(i_t)}_{\begin{smallmatrix}\textrm{average solver}\\\textrm{wallclock incurred}\end{smallmatrix}}
	\qquad=\qquad\underbrace{\min_{i\in[d]}\frac1T\sum_{t=1}^T\cost_t(i)}_{\begin{smallmatrix}\textrm{average wallclock of}\\\textrm{the best configuration}\end{smallmatrix}}
	\qquad+\qquad\underbrace{\frac{\Regret_T}T}_{\begin{smallmatrix}\\\\\textrm{vanishes as }\\T\to\infty\end{smallmatrix}}
\end{equation}
Thus, so long as the simulation has enough timesteps (large $T$), integrating an online learner with sublinear regret into a numerical simulation code will lead to performance nearly as good as that of the best fixed solver configuration.
As we will formalize in the next section, there is typically a tradeoff between the first and second right-hand terms in Equation~\ref{eq:wallclock}, as we can improve the former by considering more configurations, but doing so incurs a higher exploration cost that shows up in the regret.

As shown in the pseudo-code in Algorithm~\ref{alg:bandit}, our specific setup is a challenging type of online learning known as the {\bf adversarial bandit} setting~\citep{auer2002exp3}.
The {\em bandit} setting~\citep{bubeck2012regret} assumes access only to the value of the cost function $\cost_t$ at the selected configuration $i_t$, and not at any other $i\in[d]\backslash\{i_t\}$.
This contrasts with the {\em full information} setting~\citep{shalev-shwartz2011oco}, in which the value of $\cost_t$ is observed at every point in its domain;
doing so in a numerical simulation code would require prohibitively many solves.
Furthermore, our setting is {\bf adversarial} because the linear systems do not arise from a stationary distribution and so the cost functions do not fully depend on i.i.d. random variables, an assumption made by well-known {\em stochastic} bandit methods such as UCB~\citep{auer2002ucb} and Thompson sampling~\citep{thompson1933sampling};
in Figures~\ref{fig:frob} and~\ref{fig:costs} we validate this non-stationarity empirically.

For simplicity, we omit two details from our setup. 
The first is that the sequence of linear systems is {\em not} oblivious to the choice of solver, as in practice both $\*A_{t+1}$ and $\*b_{t+1}$ often depend on the solution returned for the system $\*A_t\*x=\*b_t$, and different solvers return different vectors in the $\varepsilon$-ball around the true solution.
While this has theoretical implications in online learning, in practice the physical state of a well-specified numerical simulation should not depend strongly on the choice of solver;
more concisely, we expect the realized sequence of linear systems to be $\BigO(\varepsilon)$-close to an oblivious sequence.\footnote{\revision{In practice this closeness depends strongly on the systems' conditioning, cf. Figure~\ref{fig:diff}.}}\looseness-1

The second detail is that an ideal configuration policy will make use of metadata about each linear system, such as what equation (\revision{pressure}, potential, etc.) is being solved, time-dependent physical quantities, and numerical information such as the initial residual.
For example, \citet{khodak2024learning} use contextual bandits to learn a configuration policy that accounts for a time-dependent diffusivity constant when solving the heat equation.
However, in this study we found that running a separate online learner for each type of equation was sufficient.

\subsection{Learning algorithm}\label{sec:algorithm}

\begin{algorithm}[!t]
	\begin{minipage}{\linewidth}
		\caption{\label{alg:pcgbandit}
			Our PCGBandit algorithm for adaptively configuring solvers across a sequence of $T$ linear system instances.
			The method's update rule differs from Tsallis-INF, as we normalize the cost estimates by the average wallclock $W/t$ observed so far.
			The optimization algorithm used to obtain the updated distribution $\*p_{t+1}$ is Newton's method initialized at $\*p_t$.
		}
	\end{minipage}
	\begin{minipage}{0.95\linewidth}
		\DontPrintSemicolon
		\KwIn{solver configuration indices $1,\dots,d$}
		$\*c\gets\*0_d$\tcp*{initialize cumulative cost estimates}
		$\*s\gets\*0_d$\tcp*{initialize variance-reduction shifts}
		$W\gets0$\tcp*{initialize wallclock tracker}
		$\*p_1\gets\*1_d/d$\tcp*{set initial distribution over configurations}
		\For{timestep $t=1,\dots,T$}{
			sample $i_t\sim\*p_t$\tcp*{pick solver configuration}
			use $i_t$ to solve $\*A_t\*x_t=\*b_t$\tcp*{solve linear system}
			observe $\cost_t(i_t)$\tcp*{get solver cost}
			$\*c_{[i_t]}\gets\*c_{[i_t]}+\frac{\cost_t(i_t)}{\*p_{t[i_t]}}$\tcp*{update cumulative cost estimates}
			\For{index $i=1,\dots,d$}{
				\If{$\*p_{t[i]}\ge16/t$}{
					$\*s_{[i]}\gets\*s_{[i]}+\frac12$\tcp*{update variance-reduction shifts}
					\If{$i=i_t$}{$\*s_{[i]}\gets\*s_{[i]}-\frac1{2\*p_{t[i]}}$}
				}
			}
			$W\gets W+\cost_t(i_t)$\tcp*{update wallclock tracker}
			$\*p_{t+1}\gets\argmin\limits_{\*p\in\triangle_d}~\langle\*p,\frac{\*c}{W/t}+\*s\rangle-\sum\limits_{i=1}^d\sqrt{t\*p_{[i]}}$\tcp*{update distribution}
		}
	\end{minipage}
\end{algorithm}

We now turn to the specific learning algorithm underlying PCGBandit.
Our method adapts the worst-case optimal learner, Tsallis-INF~\citep{abernethy2015fighting,zimmert2021tsallis}, which like most bandit algorithms works by (1)~sampling $i_t$ from a distribution $\*p_t\in\triangle_d$ over the $d$ configurations and (2)~using the feedback $\cost_t(i_t)$ to update to the next distribution $\*p_{t+1}$.\footnote{Here $\triangle_d=\{\*p\in\R_{\ge0}^d$ s.t. $\|\*p\|_1=1\}$ denotes the probability simplex over $d$ outcomes.}
Tsallis-INF incorporates the feedback into a cost estimator $\*c_t=\frac{\cost_t(i_t)}{\*p_{[i_t]}}\*e_{i_t}$ and for some learning rate $\eta_t>0$ sets the next distribution to be the minimizer
\begin{equation}\label{eq:update}
	\*p_{t+1}=\argmin_{\*p\in\triangle_d}\sum_{s=1}^t\langle\*p,\*c_s\rangle-\frac4{\eta_t}\sum_{i=1}^d\sqrt{\*p_{[i]}}
\end{equation}
of the empirical cost estimate~(first term) regularized by the negative Tsallis entropy~(second term).
This minimization can be done using Newton's method at a computational cost of $\BigO(d)$~\citep[Algorithm~2]{zimmert2021tsallis};
notably this means the learning cost does {\em not} scale with the problem size $n$.\footnote{\revision{The cost of communicating the chosen configuration $i_t$ does scale with the number of processes but is small relative to the communication cost of a typical Krylov solve.}}

If $\eta_t=\frac2{\sqrt t}$ and the cost of every configuration is bounded, then Tsallis-INF guarantees (in expectation) that $\Regret_T=\BigO(\sqrt{dT})$~\citep[Theorem~1]{zimmert2021tsallis}, which is worst-case optimal in both the number of timesteps $T$ and the number of configurations $d$.\footnote{See \citet{khodak2024learning} for the case where the configuration space is an interval of $\R$.}
Substituting this into Equation~\ref{eq:wallclock} suggests that if $T\gg d$ then our wallclock will be almost as good as if we had always used the best fixed solver.
It also suggests a tradeoff between the number of configurations $d$ we can consider and the regret;
in particular, more configurations will decrease the first right-hand term in Equation~\ref{eq:wallclock} characterizing the simulation cost, because we are minimizing over more options, but it will increase the bound of $\BigO(\sqrt{dT})$ on the regret in the second term.
We explore this tradeoff empirically in Section~\ref{sec:configs}.

We make two changes to the above for our final learning method, which is specified in Algorithm~\ref{alg:pcgbandit}.
First, unlike \citet{khodak2024learning}, we use reduced-variance~(RV) cost estimation~\citep{zimmert2021tsallis}, implemented by setting $\eta_t=\frac4{\sqrt t}$ and adding a shift $\*s\in\R^d$ to $\*c$ in the objective~\eqref{eq:update};
this is to mitigate instability due to the division by $\*p_{t[i]}$ in the default importance-weighted~(IW) cost estimates.
Second, we do not have a bound on the cost of every configuration,\footnote{This issue was formally studied by \citet{gagliolo2011algorithm}.} so we normalize the IW cost estimator $\*c$ in the update objective~\eqref{eq:update} by the average cost $\frac1t\sum_{s=1}^t\cost_s(i_s)$ seen so far.\looseness-1

Tsallis-INF is just one of many bandit algorithms that could be applied to this problem.
While our description above was motivated by worst-case optimal guarantees, such adversarial regret bounds are often viewed as pessimistic in practical non-stochastic settings that are not (colloquially)~adversarial.
However, as shown by \citet{zimmert2021tsallis}, Tsallis-INF also enjoys ``best-of-both-worlds'' guarantees, in that it simultaneously recovers the optimal stochastic-setting regret;
thus we might hope that it does well in non-stationary but still benign settings too.
Future work may consider other cost estimators, alternative ``best-of-both-worlds'' methods~\citep{xu2023bayesian}, or bandit algorithms that provably adapt to unbounded configuration costs~\citep{putta2022scale}.\looseness-1

\subsection{Preconditioner configurations}\label{sec:configurations}

We now specify the set of configurations that we learn across.
The main SPD preconditioners used in OpenFOAM are diagonal-based incomplete Cholesky~(IC(0)) and geometric agglomerated algebraic multigrid~(multigrid).
The former is a no-fill incomplete Cholesky factorization computed using a recurrence for its diagonal elements~\citep{saad2003iterative} and has
no configurable parameters.
The latter, on the other hand,  has numerous influential hyperparameters, of which we focus on which smoother to use (either Gauss-Seidel, IC(0), IC(0)+Gauss-Seidel, or symmetric Gauss-Seidel), the number of cells in the coarsest grid level (either 10, 100, or 1000), and whether or not to merge grid levels.
The first of these is tuned because it is required, the second because it was found to be influential by \citet{geise2023learning}, and the third because of its documented speedup potential on simple grids;\footnote{\revision{\url{https://doc.cfd.direct/openfoam/user-guide-v13/fvsolution}}} 
however, other multigrid settings could also be tuned.
We turn off agglomeration caching because it is not currently implemented to correctly handle varying multigrid configurations, although in principle it could do so.

In order to enhance the search space in the case of underperformance of multigrid, we also implement a new~(to OpenFOAM) preconditioner, ICT, which is the thresholded version of incomplete Cholesky with a tuneable drop-tolerance~\citep{brown1998ict}.
In typical incomplete Cholesky, i.e. IC(0), we only fill in the elements of the Cholesky factors of $\*A_t$ corresponding to the latter's nonzeros;
by contrast, ICT fills elements according to a threshold parameter indicating which ones to drop.
A lower threshold corresponds to more of the elements being filled in, thus typically reducing the number of iterations due to a better approximation of the Cholesky factors~(and thus a better approximation of $\*A_t^{-1}$ by the preconditioner).
This comes at the price of a higher per-iteration cost due to the larger number of nonzero elements.
While the threshold is a continuous parameter, we discretize it to choose between eight settings in the set $\{10^{-4},10^{-3.5},\dots,10^{-0.5}\}$;
note that \citet{khodak2024learning} show for a different preconditioner~(successive over-relaxation) that a fine-enough discretization can be enough to obtain optimality over the continuous domain.
Together with IC(0) and the multigrid settings from before this yields a search space of $d=33$ configurations.

\subsection{Practical considerations}\label{sec:considerations}

While the above largely specifies the PCGBandit algorithm, we close with a discussion of two additional practical considerations.
Firstly, since most software including OpenFOAM specifies a maximum number of PCG iterations, it is possible that certain configurations do not reach the required tolerance in time and thus return an insufficiently accurate solution, thus degrading simulation correctness.
To handle this we implement a backstop routine, where we rerun PCG with a default preconditioner~(IC(0)) once the maximum number is reached;\footnote{\revision{We allow the backstop enough iterations to ensure convergence to the specified tolerance.\looseness-1}}
this additional cost is counted towards the cost of the original configuration.

Secondly, we note that PCGBandit has two sources of randomness:
algorithmic randomness due to sampling $i_t\sim\*p_t$ and machine randomness due to $\cost_t(i_t)$ being the wallclock time.
While the former can be handled via a seed, the latter is inherent and can make debugging difficult due to lack of reproducibility.
However, in principle we can replace the wallclock time by a deterministic estimate that depends on the number of PCG iterations and counting the number of operations used during each;
we evaluate the performance of this in Section~\ref{sec:deterministic}.


\begin{table}[!t]
	\centering
	\renewcommand{\arraystretch}{0.94}
	\begin{tabular}{lccccc}
		\toprule
		simulation & SPD equation & $n$ & $T$ & wallclock & \#CPUs \\
		\midrule
		3D DNS & \revision{pressure} & 3.3e4 & 4.0e3 & 42\% & 1 \\
		\midrule
		Pitz-Daily & \revision{pressure} & 4.9e4 & 4.9e3 & 87\% & 1 \\
		\midrule
		Stefan & \revision{pressure} & 3.2e3 & 1.1e6 & 34\% & \multirow{2}{*}{1} \\
		problem & interface & 3.2e3 & 2.7e5 & 1\% & \\
		\midrule
		dam-break & \revision{pressure} & 1.3e5 & 2.2e4 & 68\% & 1 \\
		\midrule
		Shercliff & \revision{pressure} & 1.0e6 & 7.5e3 & 59\% & \multirow{2}{*}{16}\\
		flow & liquid metal potential & 1.0e6 & 2.5e3 & 13\% \\
		\midrule
		\multirow{2}{*}{fringing} & \revision{pressure} & 2.6e6 & 7.6e3 & 84\% \\
		\multirow{2}{*}{B-field} & liquid metal potential & 2.6e6 & 2.5e3 & 1\% & 16 \\
		& \hspace{-2.5mm}\revision{conducting} wall potential & 7.8e6 & 2.5e3 & 5\%\\
		\bottomrule
	\end{tabular}
	\vspace{-3mm}
	\caption{
		Information about the evaluation simulations.
		The ``wallclock'' column reports the percentage of total time spent on the equation when using IC(0)-preconditioned conjugate gradient.
		Only equations accounting for at least 1\% of the total wallclock are included.
	}
	\label{tab:simulations}
\end{table}

\section{Evaluation}\label{sec:eval}

\begin{figure}[!t]
	\centering
	\includegraphics[width=0.325\linewidth]{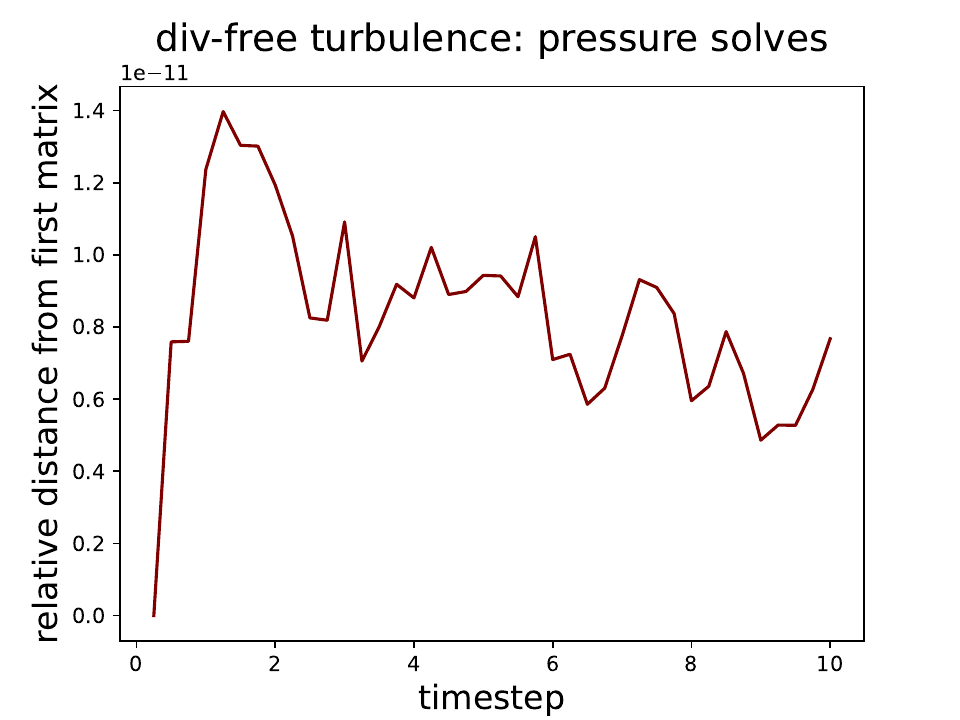}
	\includegraphics[width=0.325\linewidth]{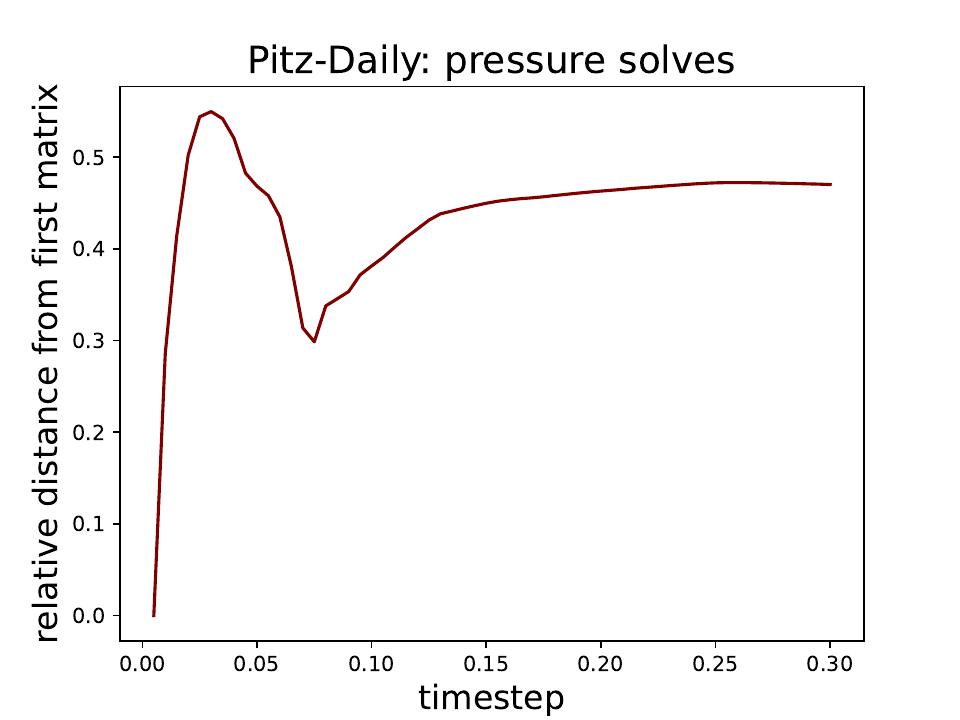}
	\includegraphics[width=0.325\linewidth]{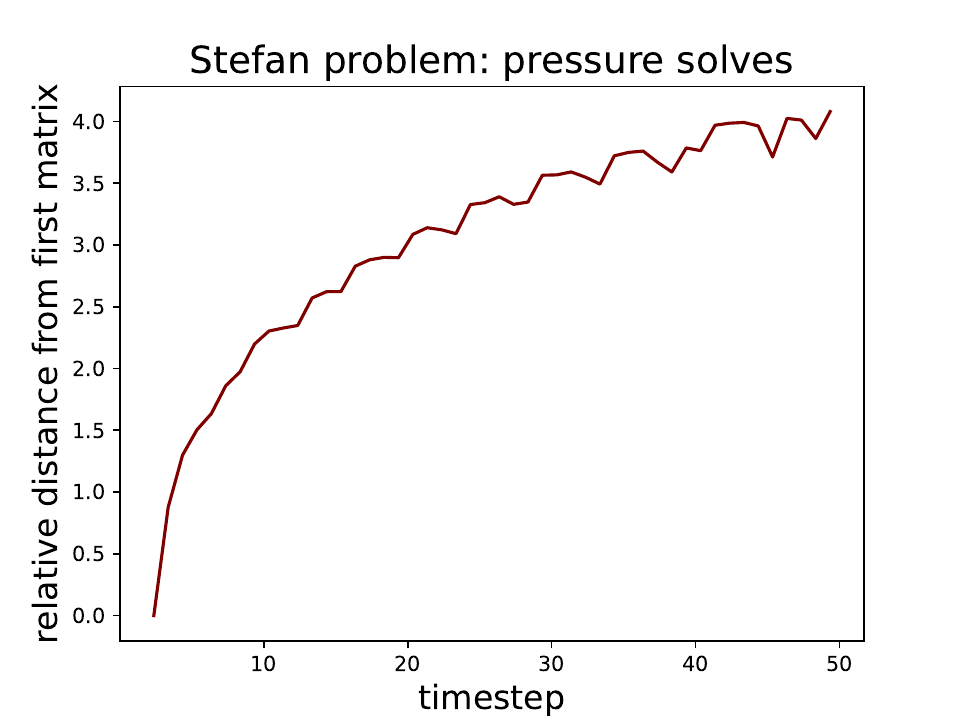}
	\includegraphics[width=0.325\linewidth]{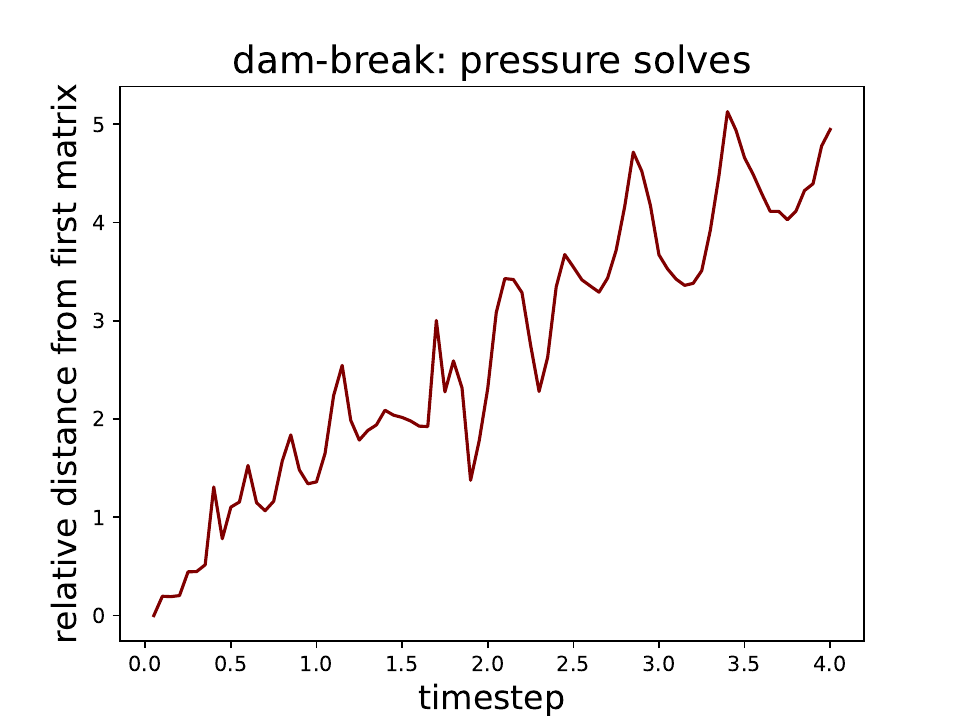}
	\includegraphics[width=0.325\linewidth]{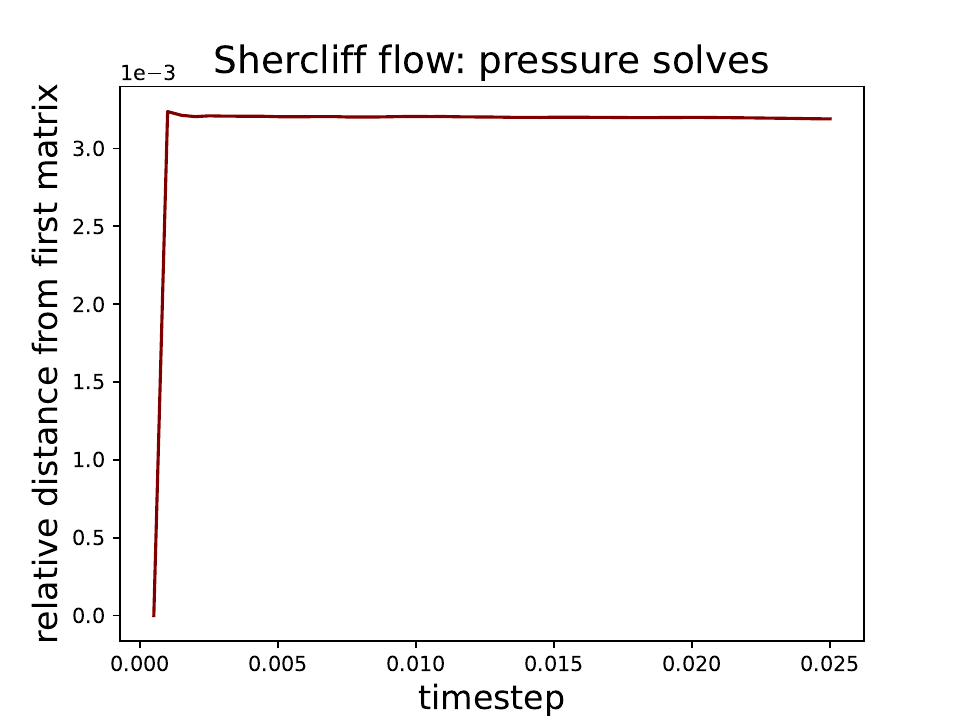}
	\includegraphics[width=0.325\linewidth]{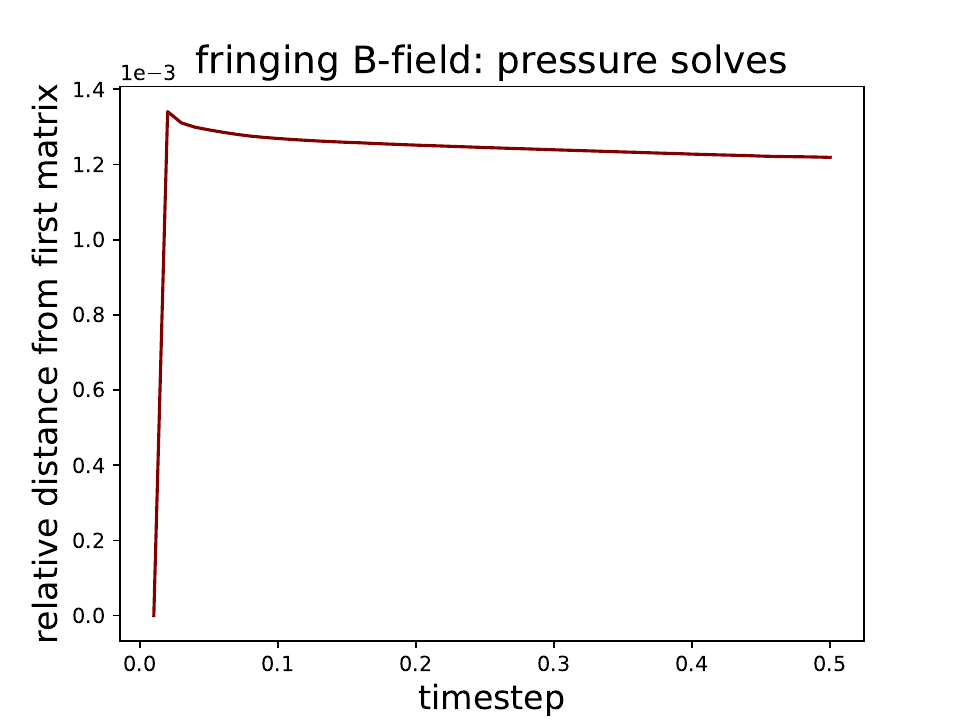}\ifdefined\arxiv\vspace{-3mm}\fi
	\caption{
		Plots of the relative variation $\|\*A_t-\*A_1\|_F/\|\*A_1\|_F$ of the linear system matrices $\*A_t$ from the initial matrix $\*A_1$ across simulation timesteps $t\in[T]$.
		Note that this is just one indicator of change, as $A_t$ can vary significantly without significant changes in $\|\*A_t-\*A_1\|_F$.
		Indeed it does not fully capture the change in difficulty of solving the systems, as Figure~\ref{fig:costs} demonstrates that the costs of the Shercliff flow and fringing B-field \revision{pressure} equations do change over time despite their matrices not changing significantly in Frobenius norm.
		By comparing with Figure~\ref{fig:plots}, we see that PCGBandit can improve simulation performance regardless of the magnitude of variation in $\|\*A_t-\*A_1\|_F$.
	}
	\label{fig:frob}
\end{figure}

\begin{figure}[!t]
	\centering
	\includegraphics[width=0.315\linewidth]{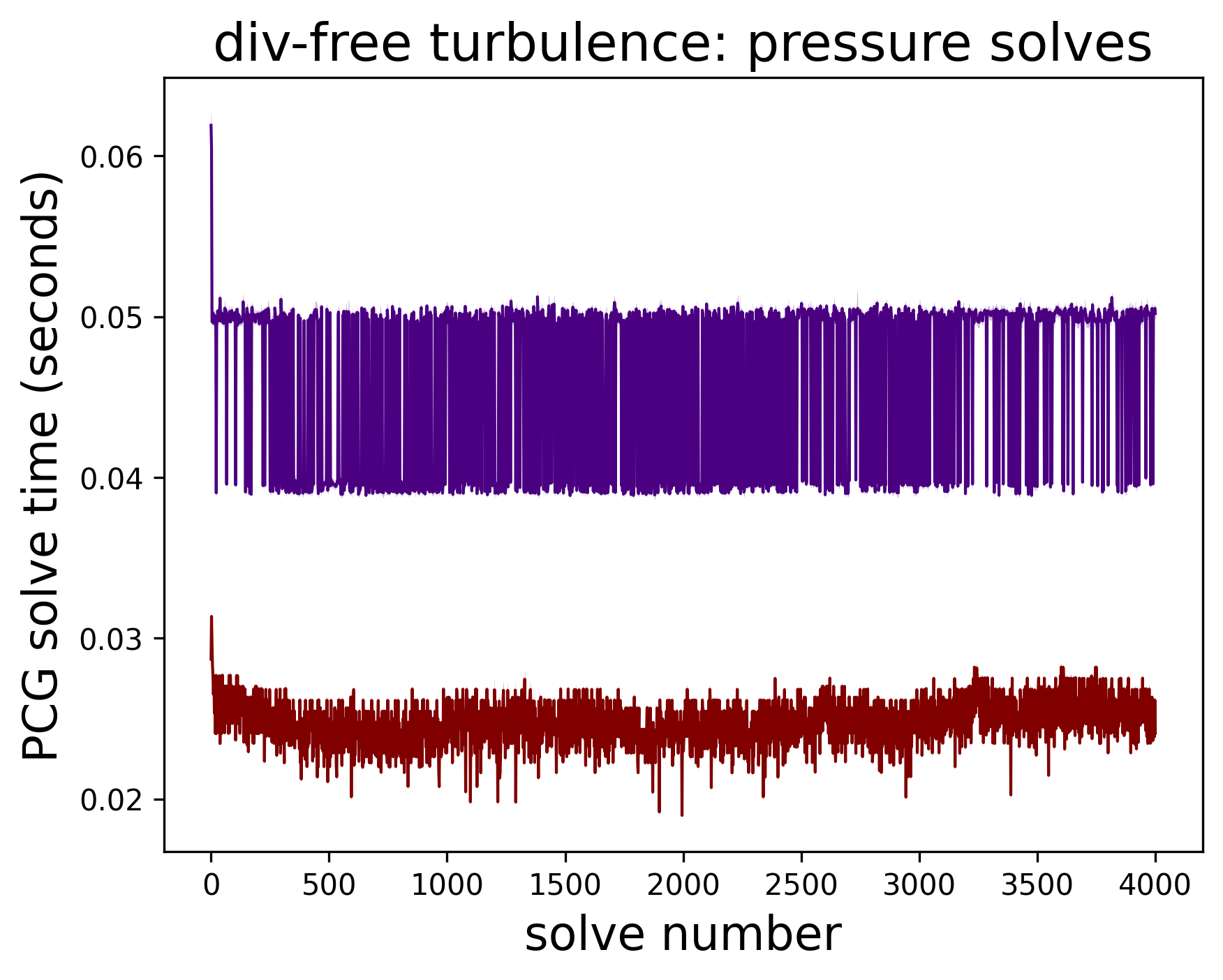}
	\includegraphics[width=0.32\linewidth]{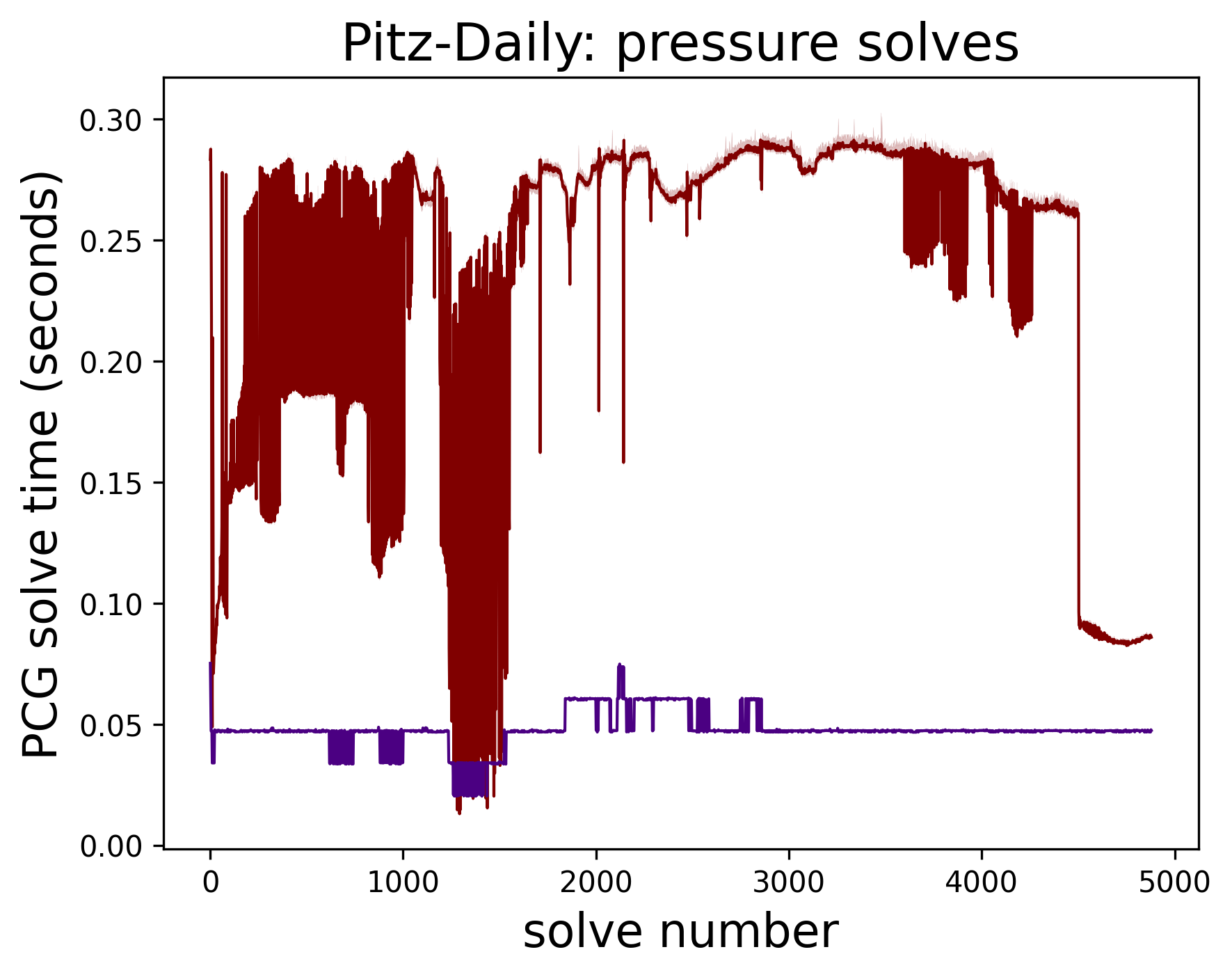}
	\includegraphics[width=0.32\linewidth]{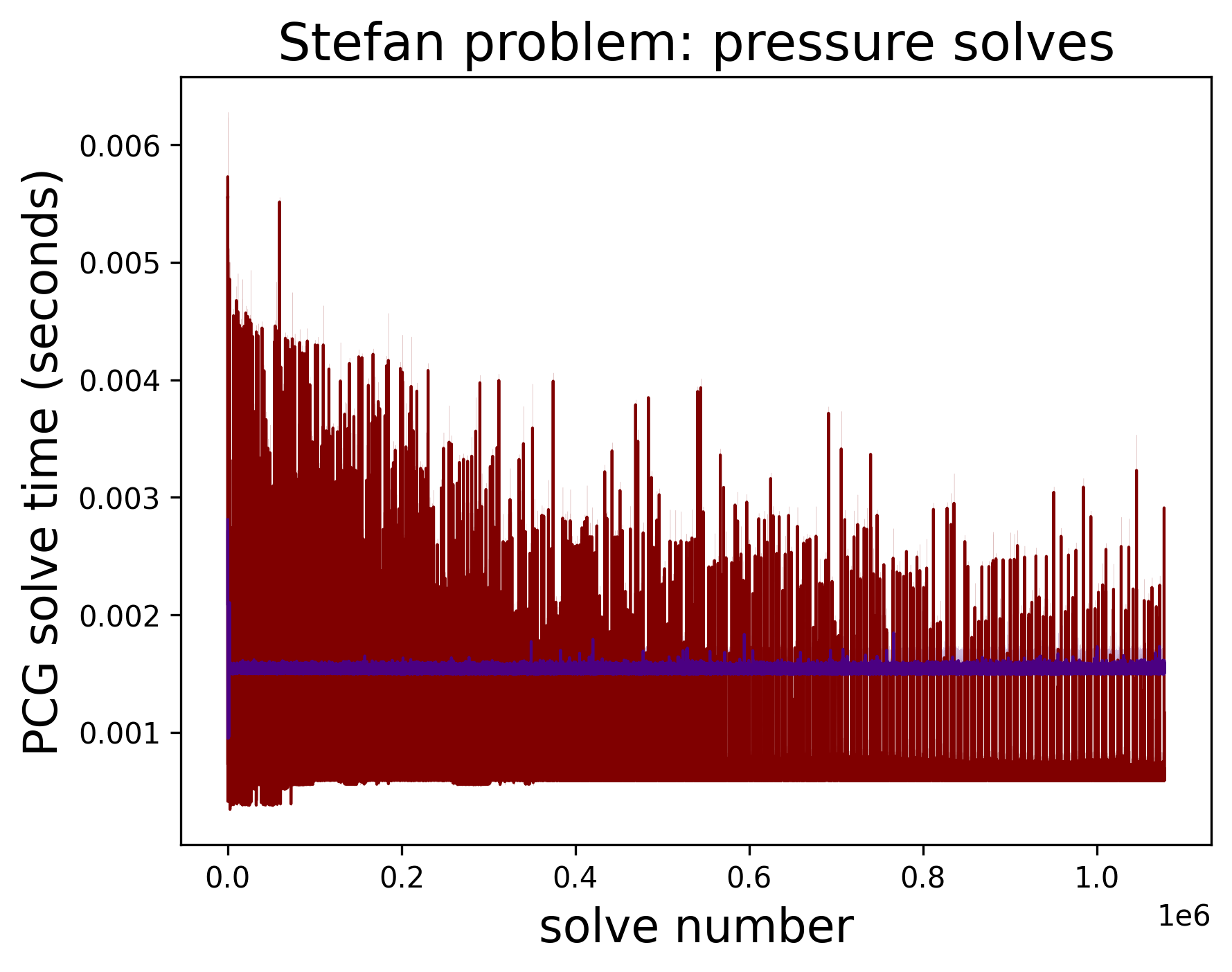}
	\includegraphics[width=0.323\linewidth]{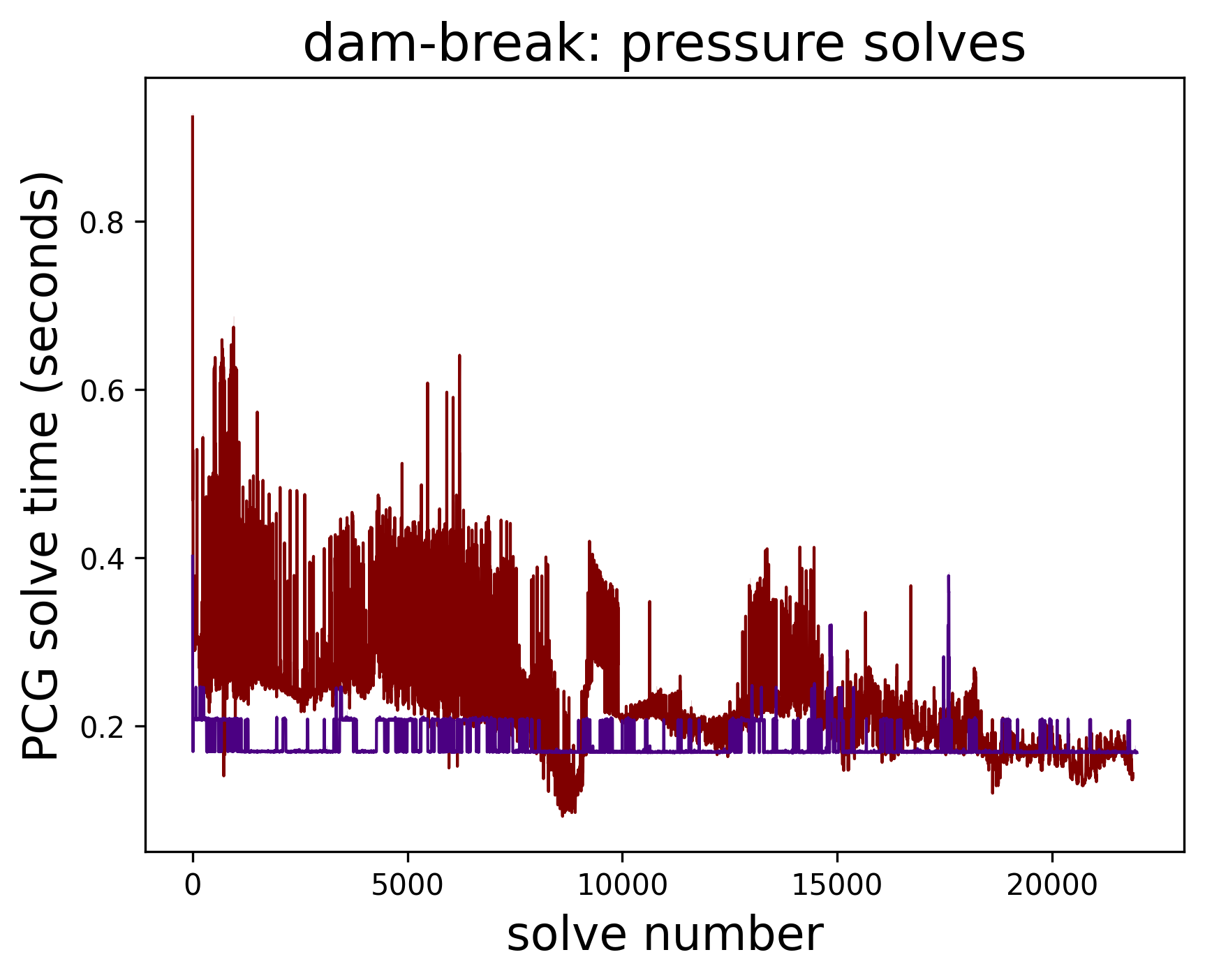}
	\includegraphics[width=0.32\linewidth]{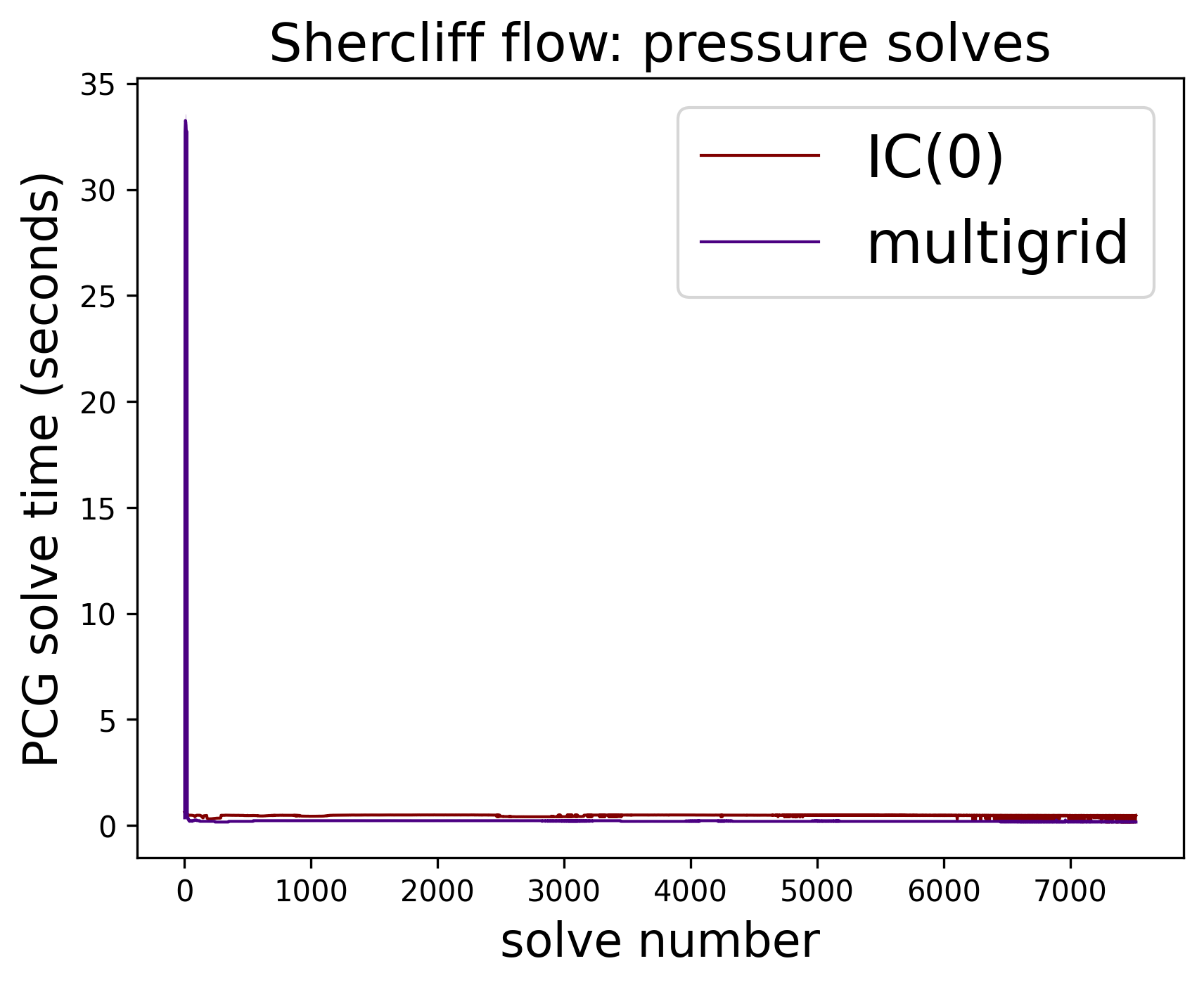}
	\includegraphics[width=0.32\linewidth]{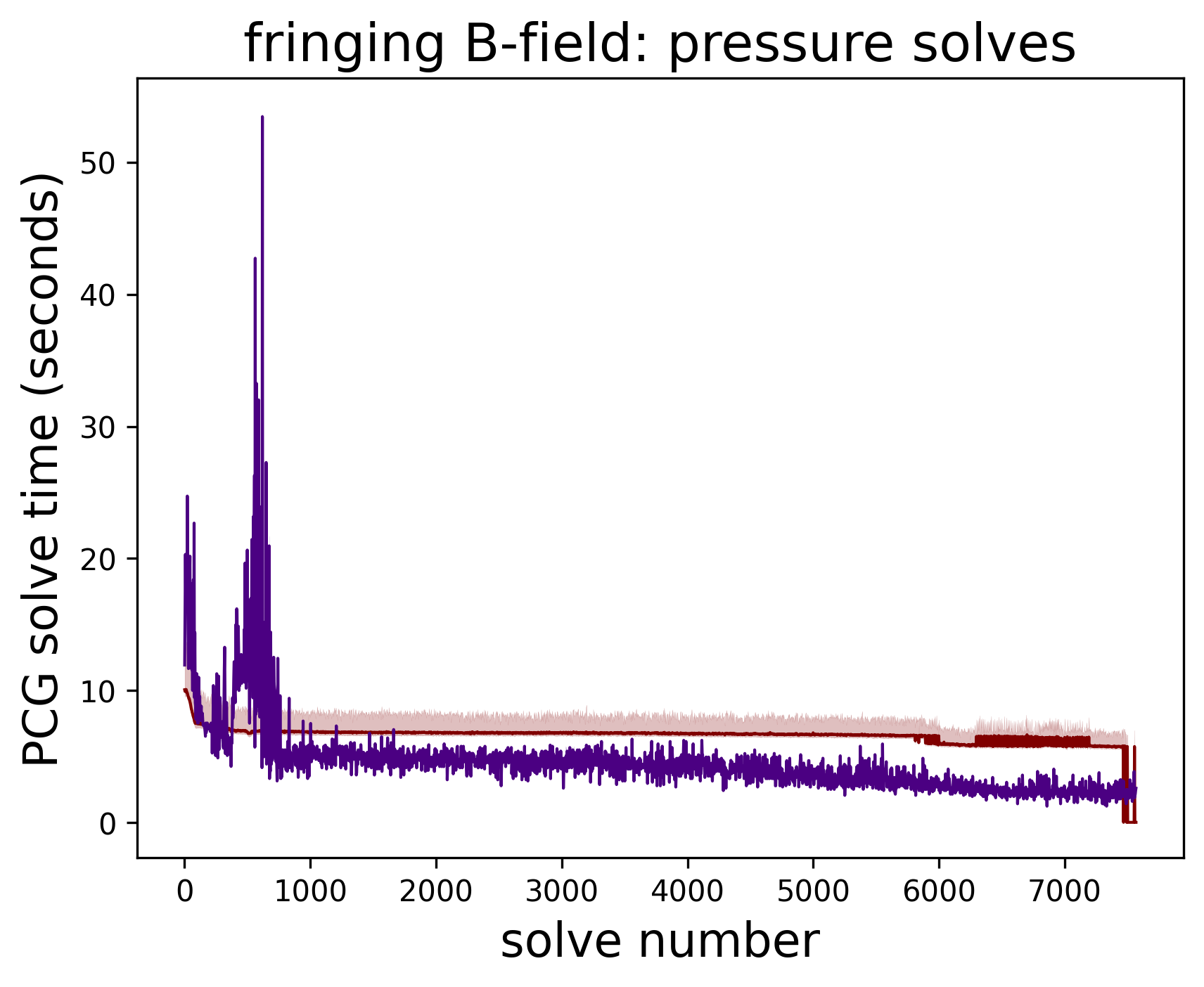}\ifdefined\arxiv\vspace{-3mm}\fi
	\caption{
		PCG solve costs using two static baseline preconditioners, IC(0) and multigrid, plotted across the timesteps of the six simulations we consider.
		These plots track the mean, minimum, and maximum costs across ten simulations, demonstrating that the trajectories are largely non-stationary, undermining the main assumption of stochastic bandit methods.\looseness-1
	}
	\label{fig:costs}
\end{figure}

We develop PCGBandit to work with the open-source numerical simulation software OpenFOAM~\citep{jasak2007openfoam}, which itself implements numerous finite volume schemes and has been built upon by other projects such as FreeMHD~\citep{wynne2025freemhd}.
To evaluate our methodology, we consider the six simulations summarized in Table~\ref{tab:simulations}, all chosen largely because linear system solving forms a large part of their wallclock cost. 
Four are OpenFOAM tutorial simulations, with their resolution doubled for added difficulty:
they comprise a 3D direct numerical simulation~(DNS) of divergence-free turbulence, a 2D Reynolds-averaged Navier-Stokes~(RANS) simulation of the Pitz-Daily backward-facing step, a multi-phase Stefan problem simulation, and a multi-fluid~(water and air) dam-break simulation.
The other two are larger-scale magnetohydrodynamics simulations taken from FreeMHD's test-cases;
they both model a 3D conductive flow, either through a closed channel with insulating walls~(Shercliff flow) or through a pipe in a fringing magnetic field~(fringing B-field);
see \citet[Section~III]{wynne2025freemhd} for further details.
We run the simulations for 25$\times$ the time it takes their magnetic fields to ramp up, which suffices to reach a steady-state.
Unlike typical OpenFOAM fluid schemes, in addition to a \revision{pressure} equation FreeMHD discretizes and solves a Poisson equation associated with the flow's electric potential~\citep[Section~II]{wynne2025freemhd}, which can account for a non-negligible part of the simulation's wallclock~(Table~\ref{tab:simulations}).
As shown in Figures~\ref{fig:frob} and~\ref{fig:costs}, the six simulations we consider have diverse types of time-dependent variation in their linear systems.\looseness-1

To assess the effectiveness of PCGBandit at accelerating OpenFOAM simulations, we compare it to running PCG with two baseline preconditioners:
IC(0) and multigrid.
For the latter we select the smoother used most often in the tutorials~(IC(0)+Gauss-Seidel) as the default,\footnote{All other multigrid options have defaults provided by OpenFOAM.} and for fair comparison with the options in the PCGBandit search space we also implement a backstop and turn off agglomeration caching.
In addition to these baselines, we compare PCGBandit to two alternative bandit methods: (1)~Tsallis-INF with importance-weighted~(IW) cost estimation, which was used by \citet{khodak2024learning}, and (2)~Thompson sampling, specifically its Gaussian variant~\citep{agrawal2017near}.
When cost permits we also compare to preconditioners drawn uniformly at random from the configuration space described in Section~\ref{sec:configurations}, as well as to the best-in-hindsight preconditioner from the same.\looseness-1

We next detail the main results, including wallclock improvements and the number of timesteps needed for PCGBandit to be useful.
Then we investigate the specific configurations found by the method, demonstrate its performance using deterministic cost estimators, and discuss the effect of search space size.\looseness-1

\subsection{Main results}

\begin{figure}[!t]
	\centering
	\includegraphics[width=0.495\linewidth]{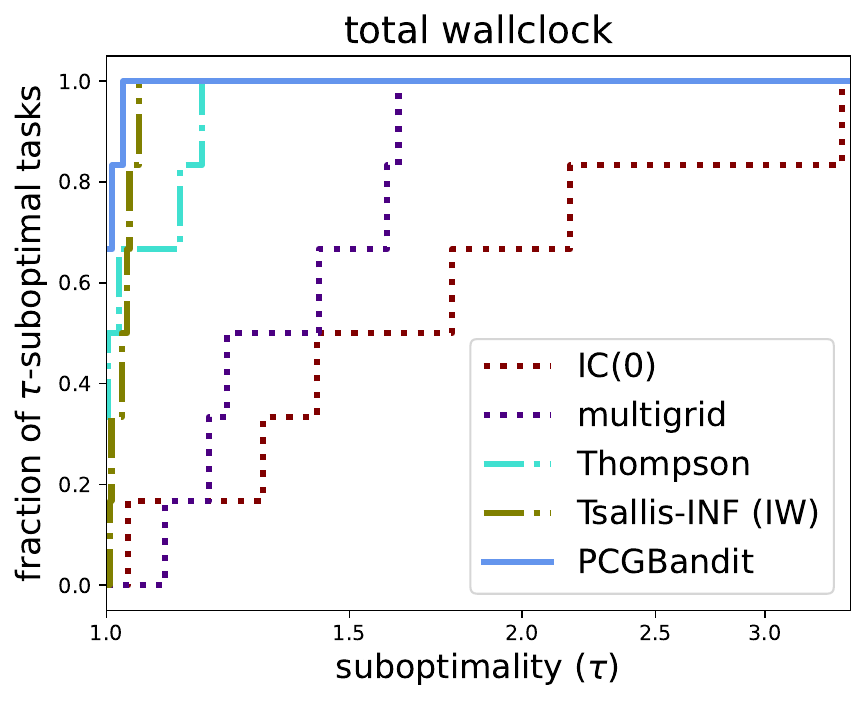}
	\includegraphics[width=0.495\linewidth]{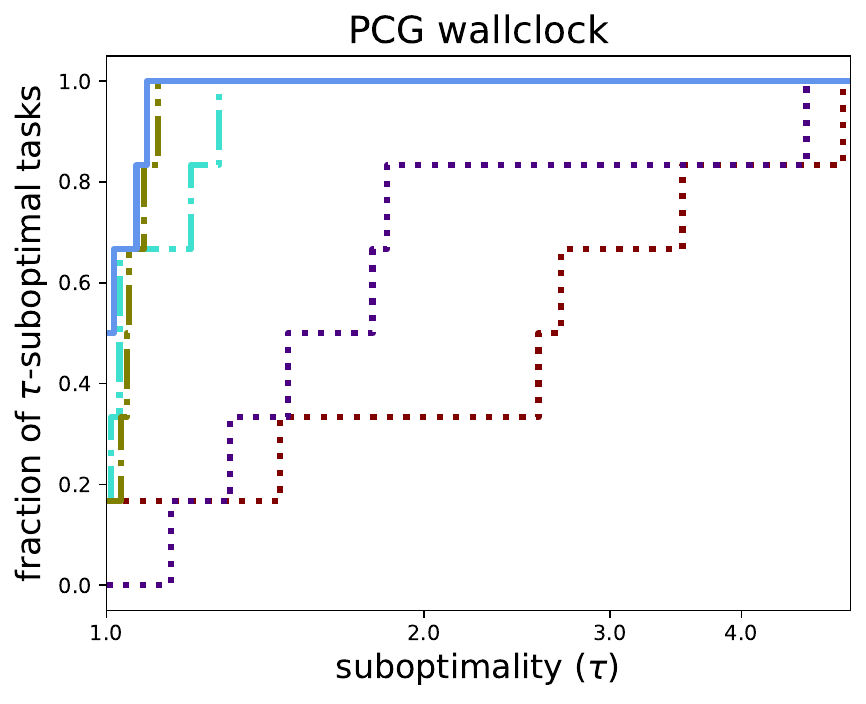}\ifdefined\arxiv\vspace{-3mm}\fi
	\caption{
		Performance profiles~\citep{dolan2002profiles} comparing bandit algorithms to two standard baselines:
		incomplete Cholesky and geometric-algebraic multigrid with the defaults described in Section~\ref{sec:eval}.
		The curves report the fraction of the six evaluation tasks from Table~\ref{tab:simulations} for which the corresponding method is $\tau$-suboptimal, i.e. has $\tau\times$ greater wallclock cost, relative to the best method of the three on each task.
		The PCGBandit~(Tsallis-INF~(RV)) curve being in the upper left implies it is optimal or near-optimal on all evaluated tasks.
	}
	\label{fig:perfprof}
\end{figure}

Each simulation is run for ten trials each on 2.8 GHz Intel Cascade Lake CPUs.
The trials measure the total wallclock, the subset of that time spent on solving SPD systems using PCG, and the subset of the latter spent on learning~(where applicable).
We start by reporting aggregate results across tasks in Table~\ref{tab:simulations} in Figure~\ref{fig:perfprof}, where we chart the performance profiles~\citep{dolan2002profiles} of the two static preconditioner baselines and three bandit methods.
As $\tau\ge1$ varies, these curves track the fraction of tasks on which each method is at most $\tau$-suboptimal.
The results demonstrate that PCGBandit~(Tsallis-INF~(RV)) is optimal or near-optimal on all tasks, with both multigrid and IC(0) being around $1.5\times$ or more slower on three tasks each in terms of total wallclock.
This effect is even more pronounced when counting only the time spent running PCG, where PCGBandit can be more than $4\times$ faster than both.
Lastly, among our choices of learning algorithms, Tsallis-INF~(IW) is only slightly worse than PCGBandit, while Thompson sampling underperforms more significantly, perhaps due to its stationarity assumption on the costs.\looseness-1

\begin{figure}[!t]
	\centering
	\includegraphics[width=0.325\linewidth]{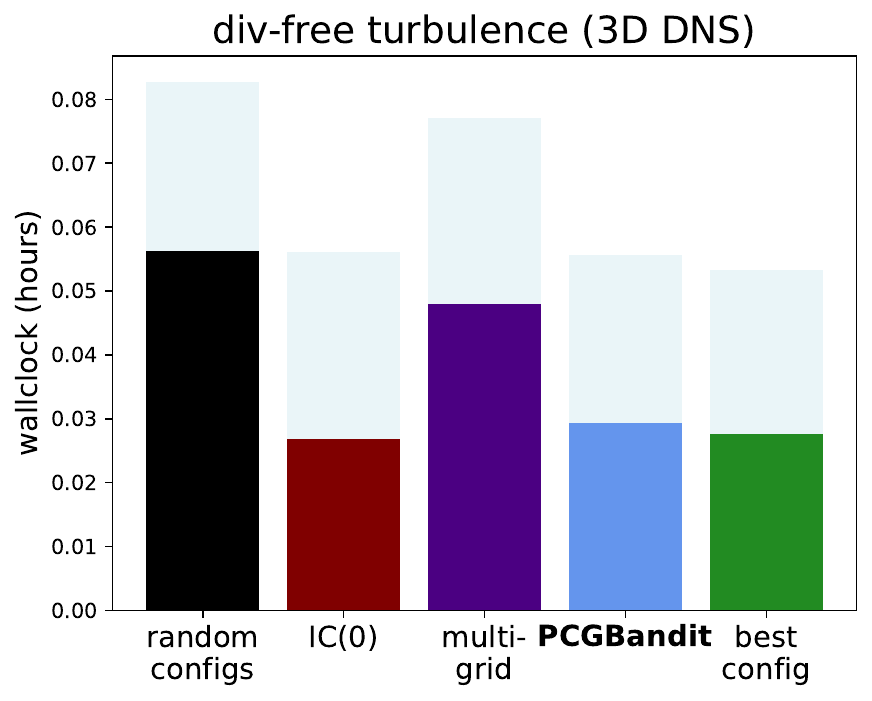}
	\includegraphics[width=0.325\linewidth]{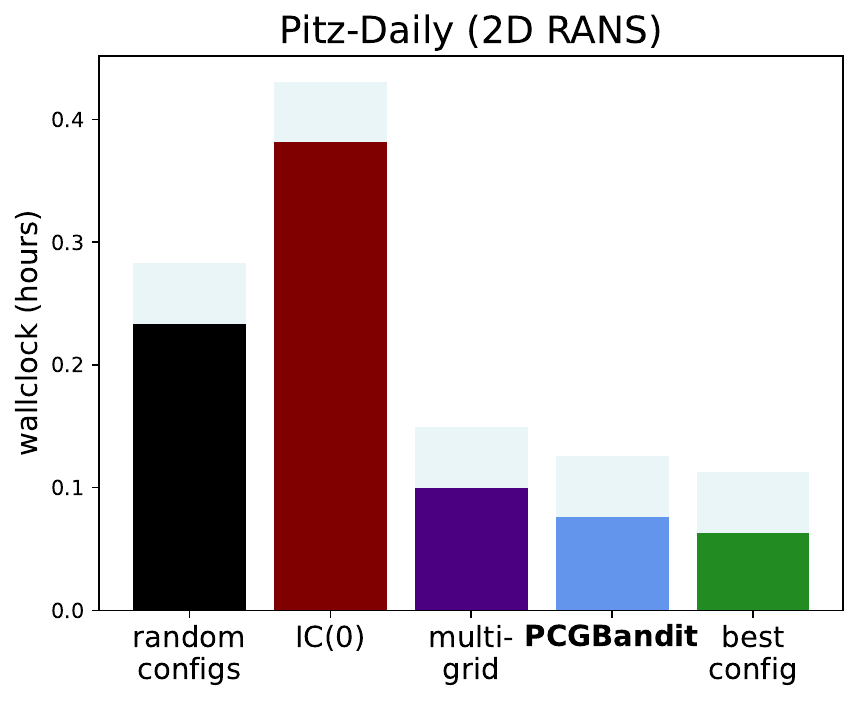}
	\includegraphics[width=0.325\linewidth]{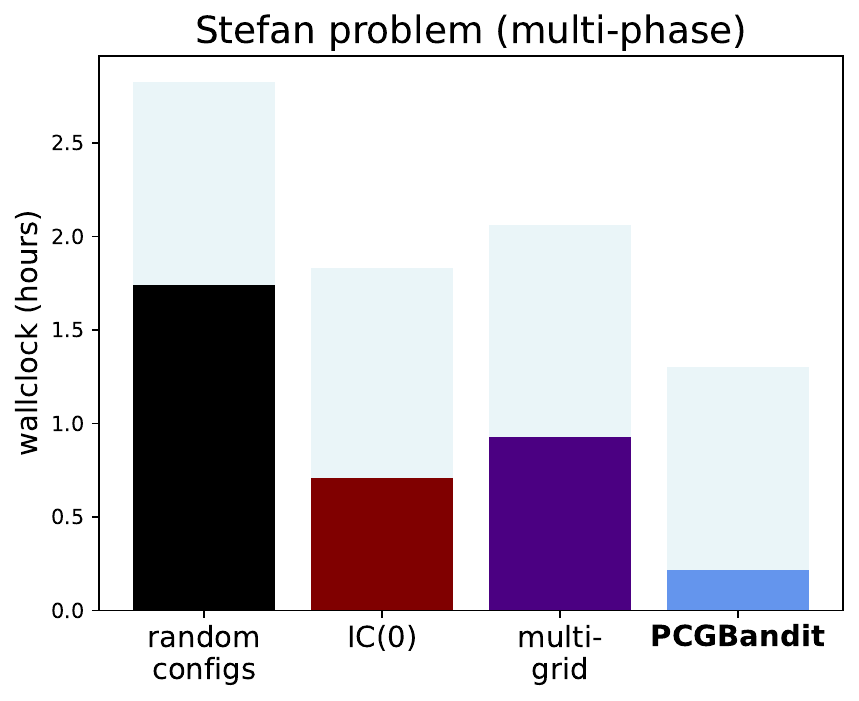}
	\includegraphics[width=0.325\linewidth]{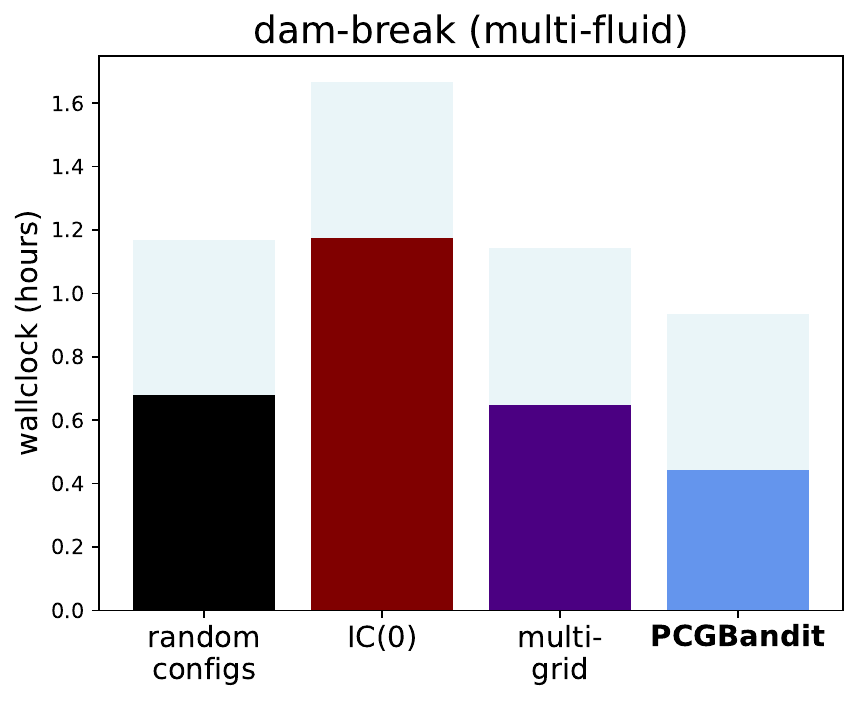}
	\includegraphics[width=0.325\linewidth]{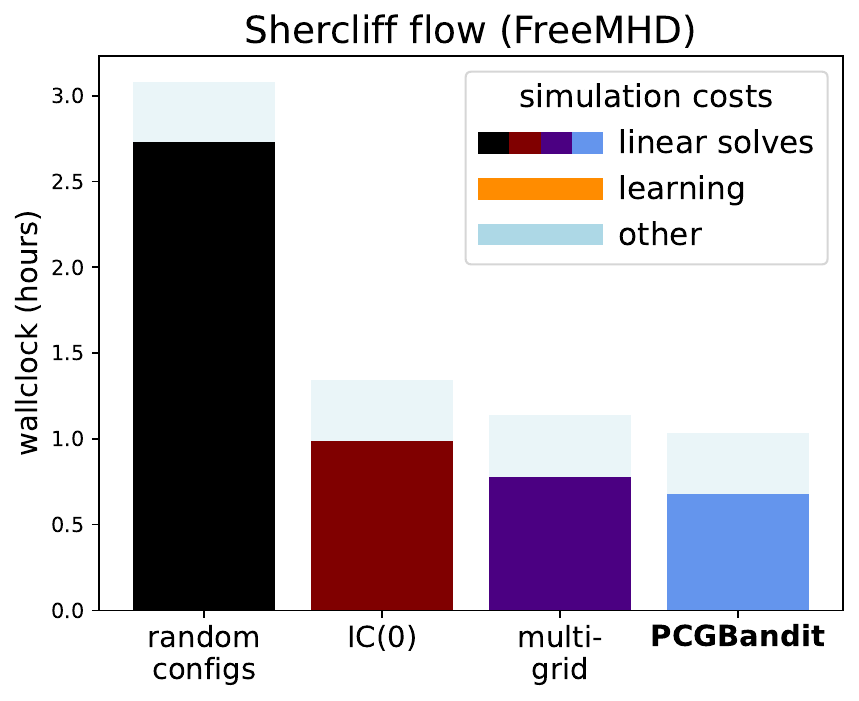}
	\includegraphics[width=0.325\linewidth]{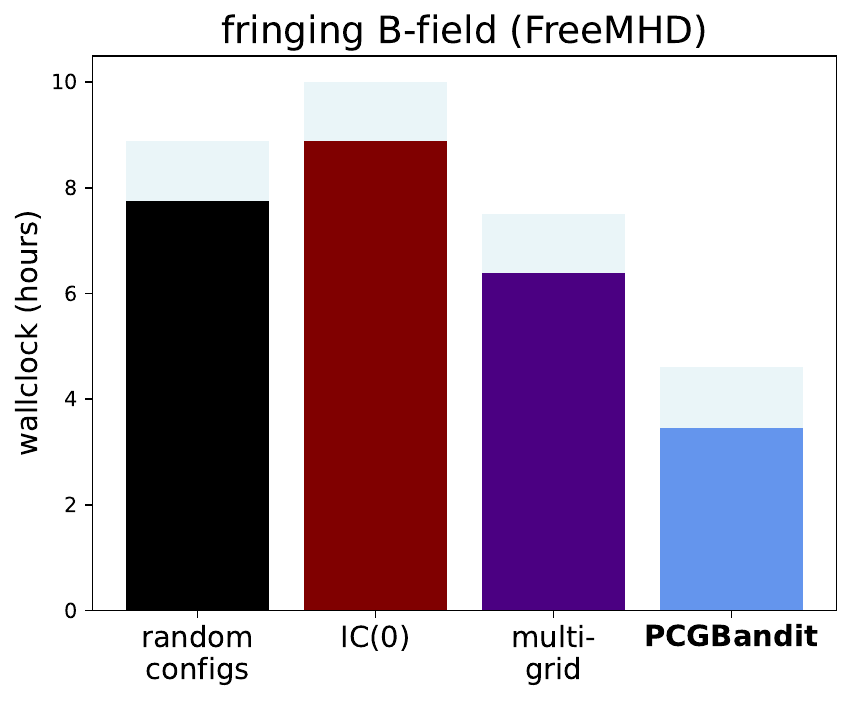}\ifdefined\arxiv\vspace{-3mm}\fi
	\caption{
		Separate wallclock comparisons of PCGBandit and baseline methods on six evaluation tasks.
		The bold component of each bar represents the time used during the simulations to solve SPD linear systems, while the remaining light component comprises all other costs.\looseness-1
	}
	\label{fig:plots}
	\vspace{5mm}
\end{figure}

We next disaggregate the same results in Figure~\ref{fig:plots}, which depicts the total wallclock as a sum of the PCG cost, the learning cost, and other costs.
Here we add the random baseline to the two static preconditioner baselines, and for the 3D DNS and Pitz-Daily simulations we show the best configuration from the PCGBandit search space.
We see that PCGBandit only fails to beat the two static baselines on the simplest task, 3D DNS, where IC(0) is near optimal.
It nevertheless comes reasonably close to the performance of the best-in-hindsight in both settings where it is known, as expected from the theory.
Lastly, note that learning costs are so negligible that they are not discernible on the graphs.\looseness-1

\begin{figure}[!t]
	\centering
	\includegraphics[width=\linewidth]{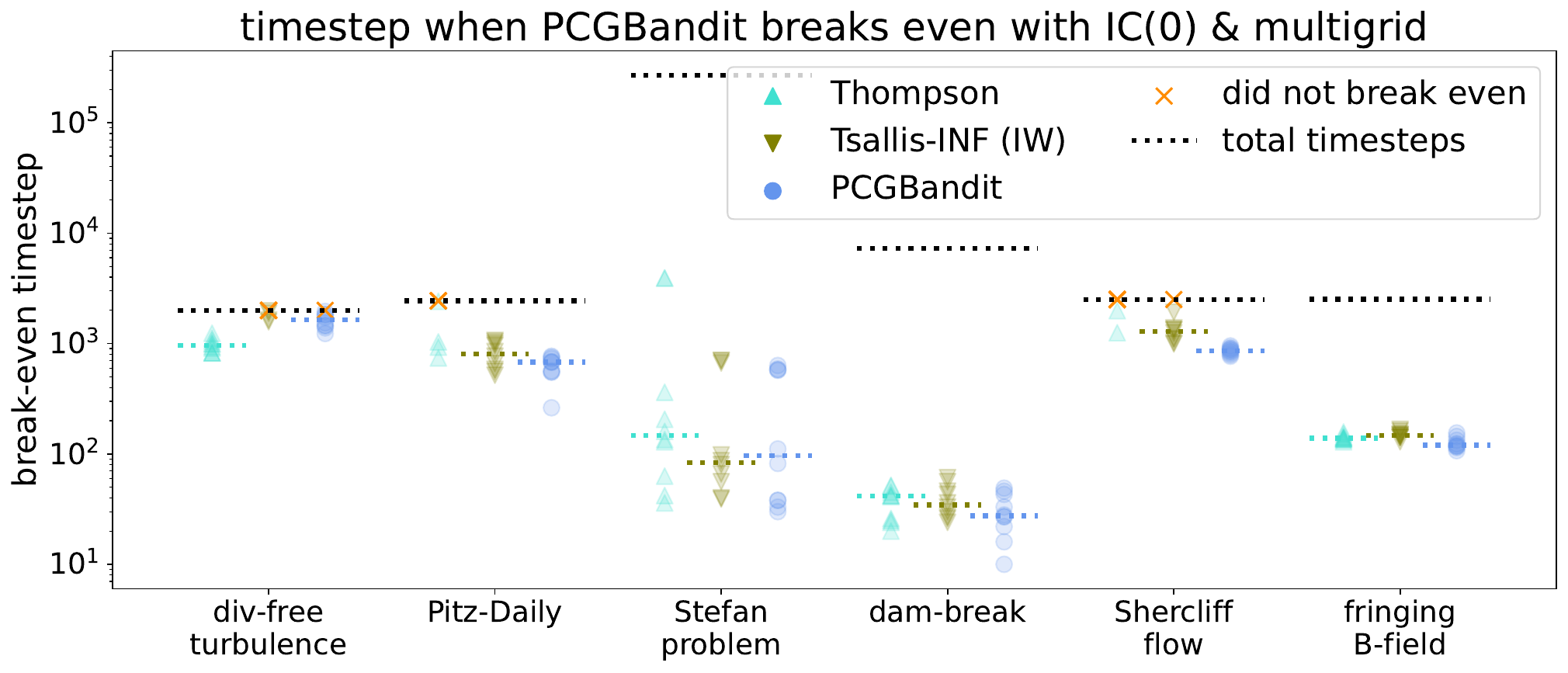}\ifdefined\arxiv\vspace{-3mm}\fi
	\caption{
		Break-even timesteps for each simulation setting, seed, and bandit algorithm.
		These timesteps are defined to be the one after which the learner's cumulative cost is smaller than that of both baselines~(IC(0) and multigrid) for the rest of the simulation.
		This indicates the minimum number of timesteps needed to be useful in a specific setting.
		The results demonstrate that PCGBandit~(Tsallis-INF~(RV)) usually breaks even around a thousand timesteps, and often needs a hundred or less.\looseness-1
	}
	\label{fig:breakeven}
\end{figure}

Finally, in Figure~\ref{fig:breakeven} we plot for each bandit algorithm the {\em break-even timestep} of each trial, defined as the timestep $t\in[T]$ after which the cumulative cost is always lower for the learner than it is for the two baselines.
Because the methods are {\em anytime}---they do not take $T$ as an input parameter---the breakeven timesteps indicate the number of linear system instances a simulation needs to see before learning is useful.
For PCGBandit~(Tsallis-INF~(RV)), one thousand steps is usually enough data, although in some cases far fewer systems are needed.
Tsallis-INF~(IW) and Thompson sampling typically need more, except the latter outperforms on the turbulence simulation~(3D DNS);
this is likely explained by the stationarity of that setting's costs~(see Figure~\ref{fig:costs}).
Overall, the chart provides some indication of what kind of simulations can benefit from learning, although as explored in Section~\ref{sec:configs} the break-even timestep can heavily depend on the number of configurations $d$.\looseness-1

\begin{figure}[!t]
	
	\begin{minipage}{\linewidth}
		\centering
		\includegraphics[width=0.491\linewidth]{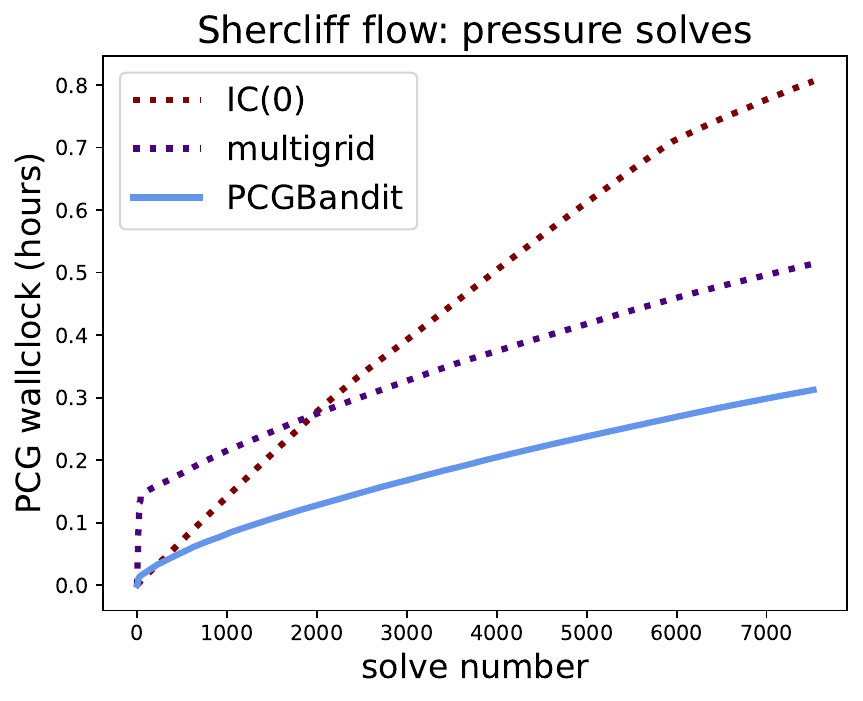}
		\includegraphics[width=0.499\linewidth]{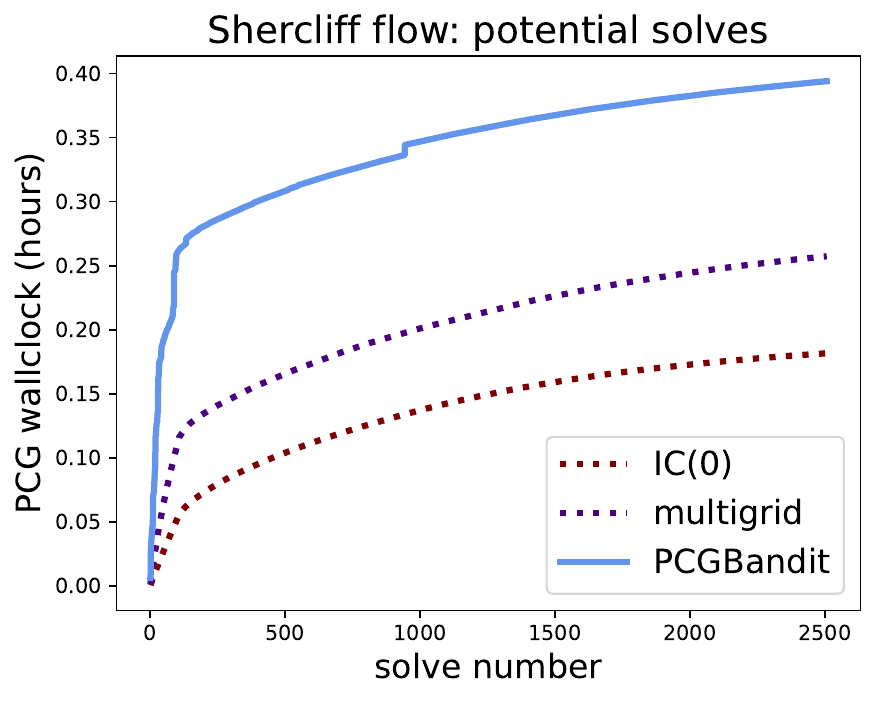}\ifdefined\arxiv\vspace{-3mm}\fi
		\caption{
			Cumulative wallclock cost of PCGBandit and the two static baseline preconditioners when solving the \revision{pressure} equation~(left) and the~(liquid metal) electric potential~(right) on the Shercliff flow MHD simulation.
		}
		\label{fig:shercliff-perf}
	\end{minipage}
	\begin{minipage}{\linewidth}
		\vspace{3mm}
		\centering
		\includegraphics[width=0.495\linewidth]{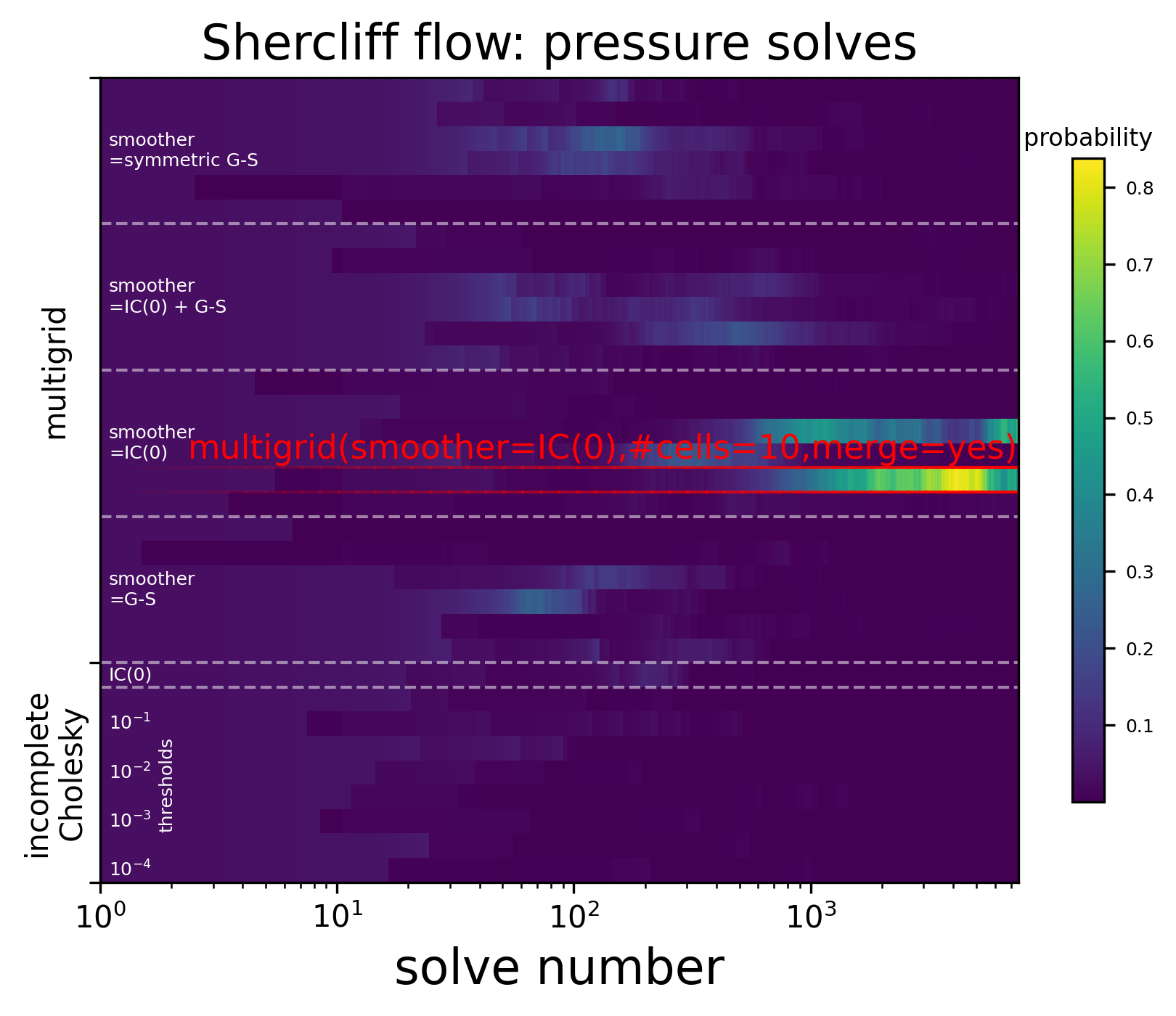}
		\includegraphics[width=0.495\linewidth]{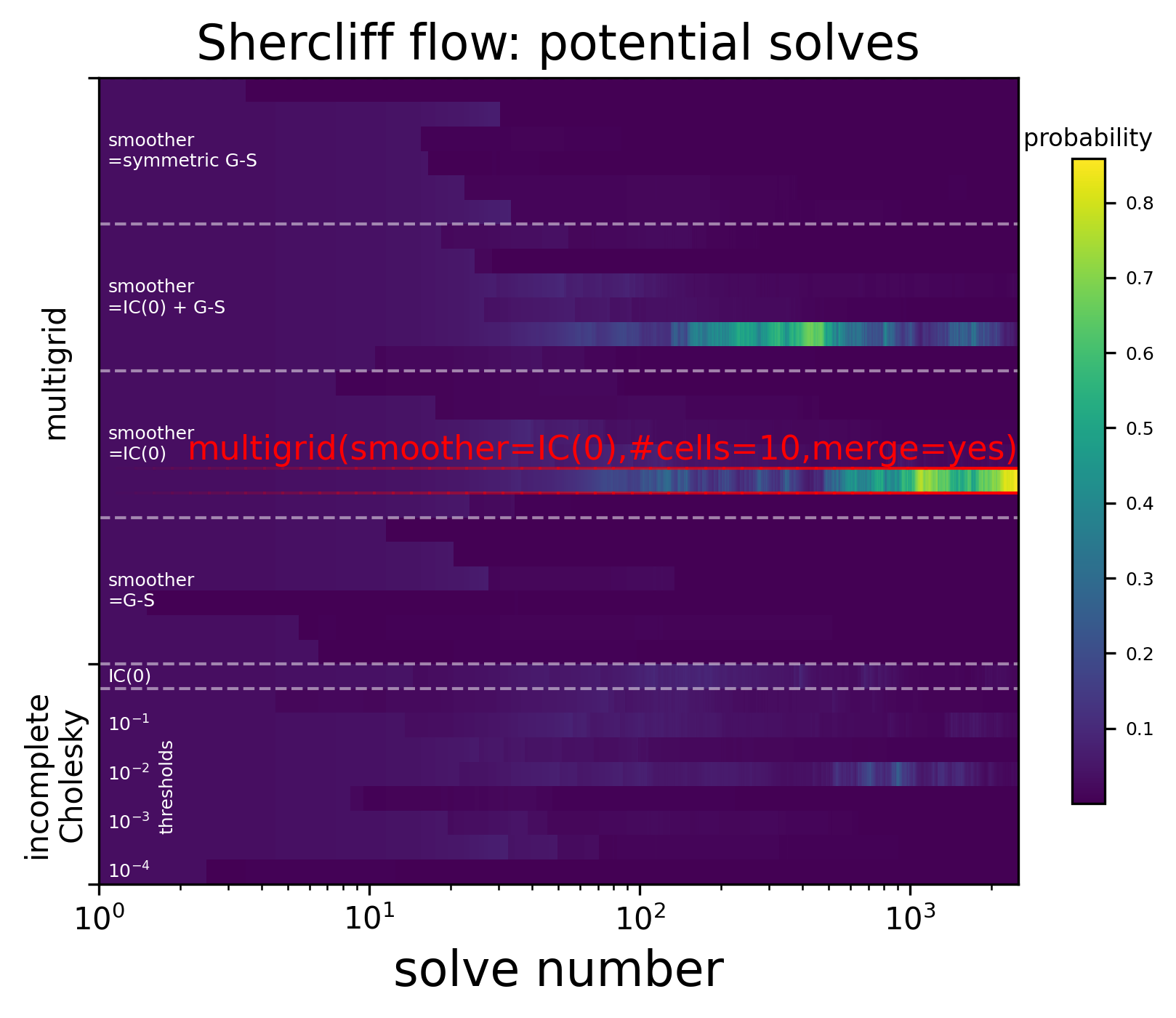}\ifdefined\arxiv\vspace{-3mm}\fi
		\caption{
			Heatmaps of probabilities assigned to each configuration at each solve of the \revision{pressure} equation~(left) and the electric potential~(right) on the Shercliff flow simulation.
			The highlighted configuration is the one with the largest total probability across solves.\looseness-1
		}
		\label{fig:shercliff-prob}
	\end{minipage}
	
\end{figure}

\subsection{A closer look at within-simulation learning}

We now take a closer look at the learning process on the two MHD simulations, Shercliff flow and fringing B-field; 
as shown in Figure~\ref{fig:plots}, the improvement due to PCGBandit is much more substantial on the latter.
In Figure~\ref{fig:shercliff-perf} we look at performance separately across the two main Shercliff flow equations and see that while PCGBandit noticeably improves the solving of the \revision{pressure} equation~(left), it performs poorly on electric potential~(right).
Much of this is due to large costs incurred very early in the simulation, while later PCGBandit's costs are similar to those of the baselines.
As shown in Figure~\ref{fig:costs}, early solves dominate the runtime on Shercliff flow, making it a hard case for a bandit algorithm since it must do well without having seen much data.
Notably, Shercliff flow was also one of the worst simulations for random sampling when comparing it to a fixed baseline~(see Figure~\ref{fig:plots}, top right), making it not surprising that a method such as PCGBandit that begins with a uniform distribution may struggle more there.
To fix this issue of trying many bad configurations, future work can consider better initial distributions or bandit algorithms that take advantage of known relationships between methods~\citep{valko2014spectral}, e.g. to use the fact that different multigrid configurations will perform more similar to each other than to different incomplete Cholesky variants.\looseness-1

\begin{figure}[!t]

\begin{minipage}{\linewidth}
	\centering
	\includegraphics[width=0.486\linewidth]{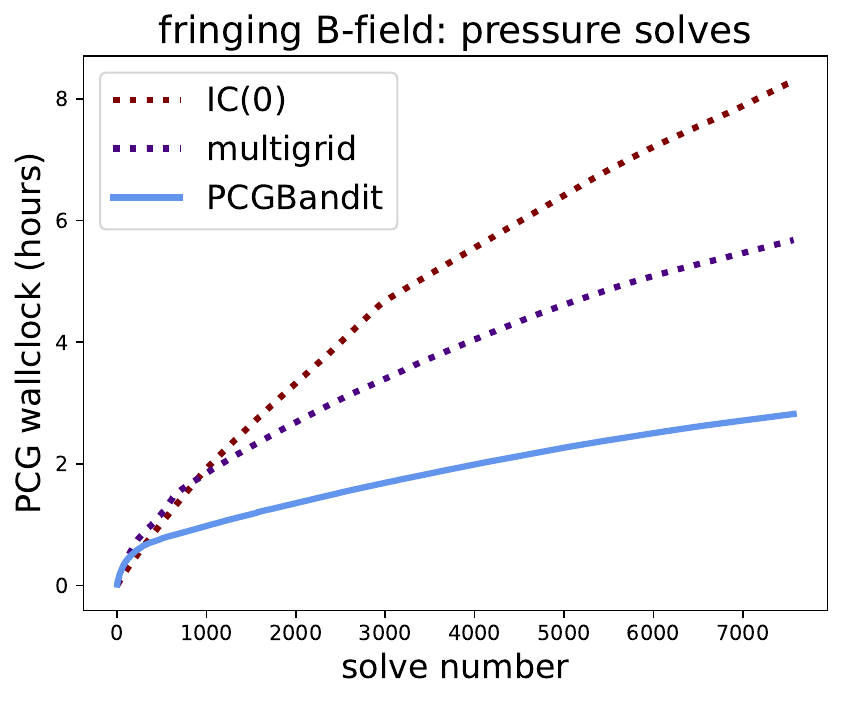}
	\includegraphics[width=0.505\linewidth]{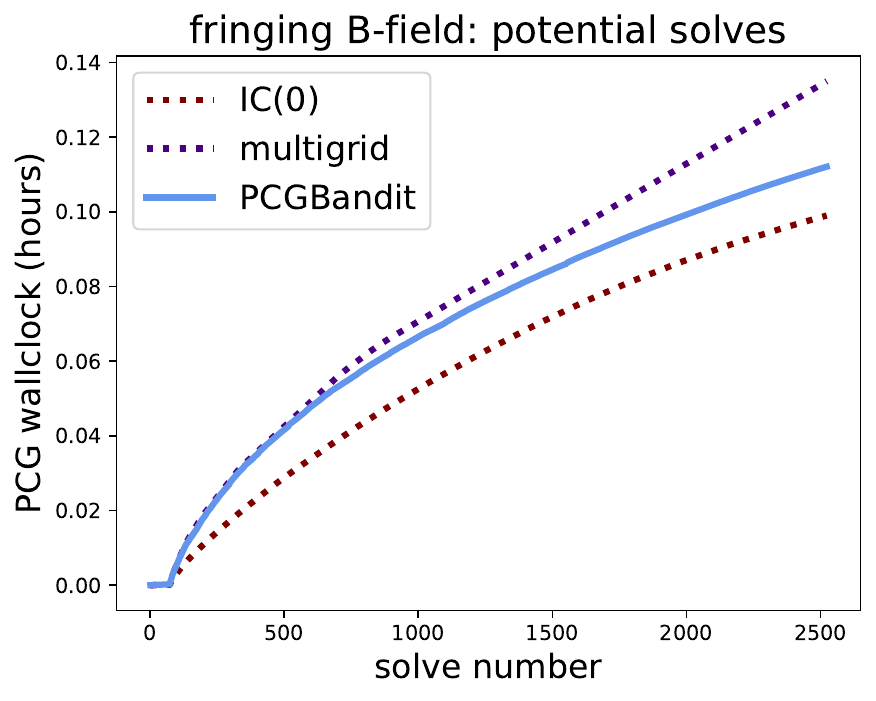}\ifdefined\arxiv\vspace{-3mm}\fi
	\caption{
		Cumulative wallclock cost of PCGBandit and the two static baseline preconditioners when solving the \revision{pressure} equation~(left) and the~(liquid metal) electric potential~(right) on the fringing B-field MHD simulation.
	}
	\label{fig:fringing-perf}
\end{minipage}

\begin{minipage}{\linewidth}
	\vspace{3mm}
	\centering
	\includegraphics[width=0.495\linewidth]{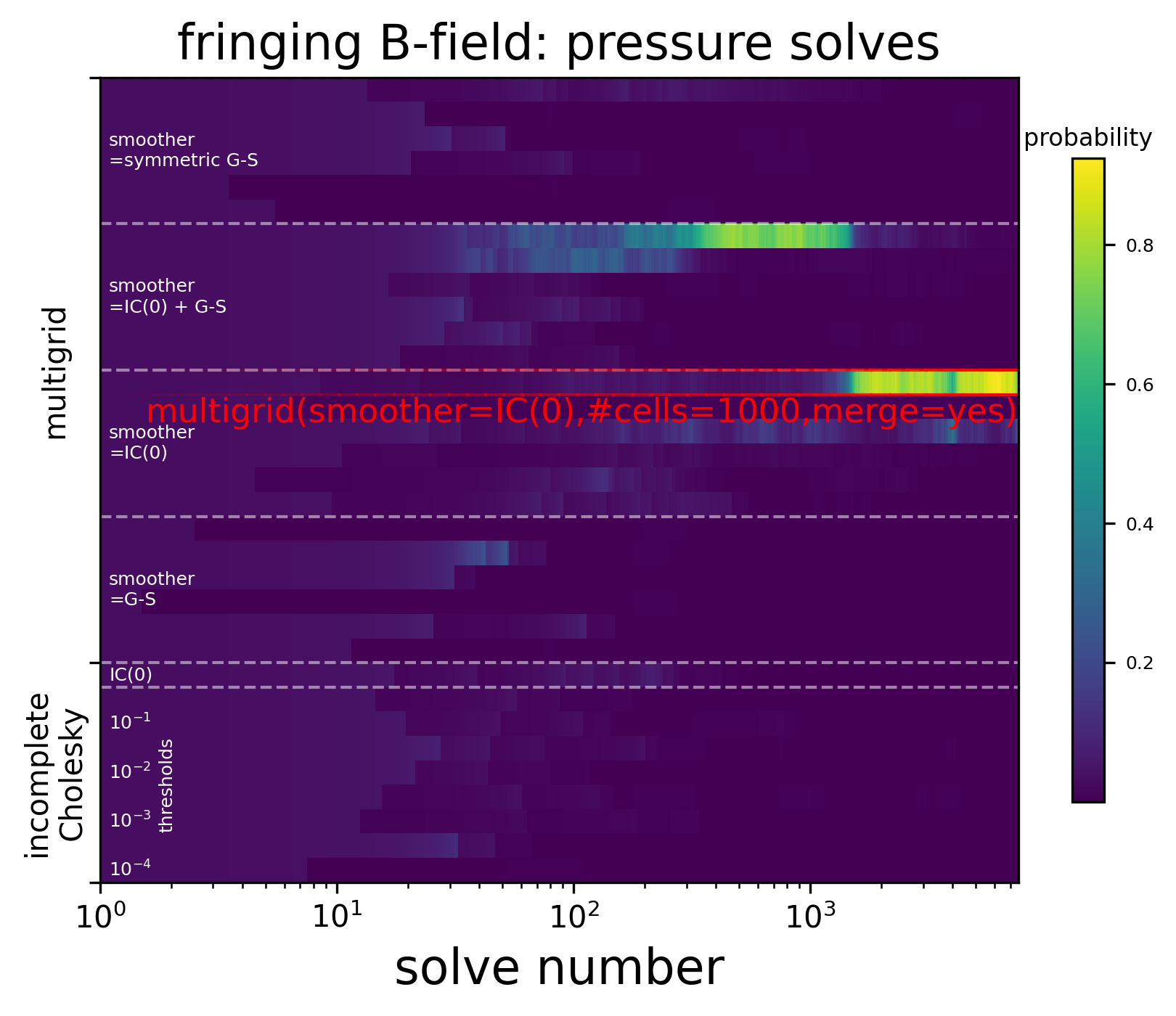}
	\includegraphics[width=0.495\linewidth]{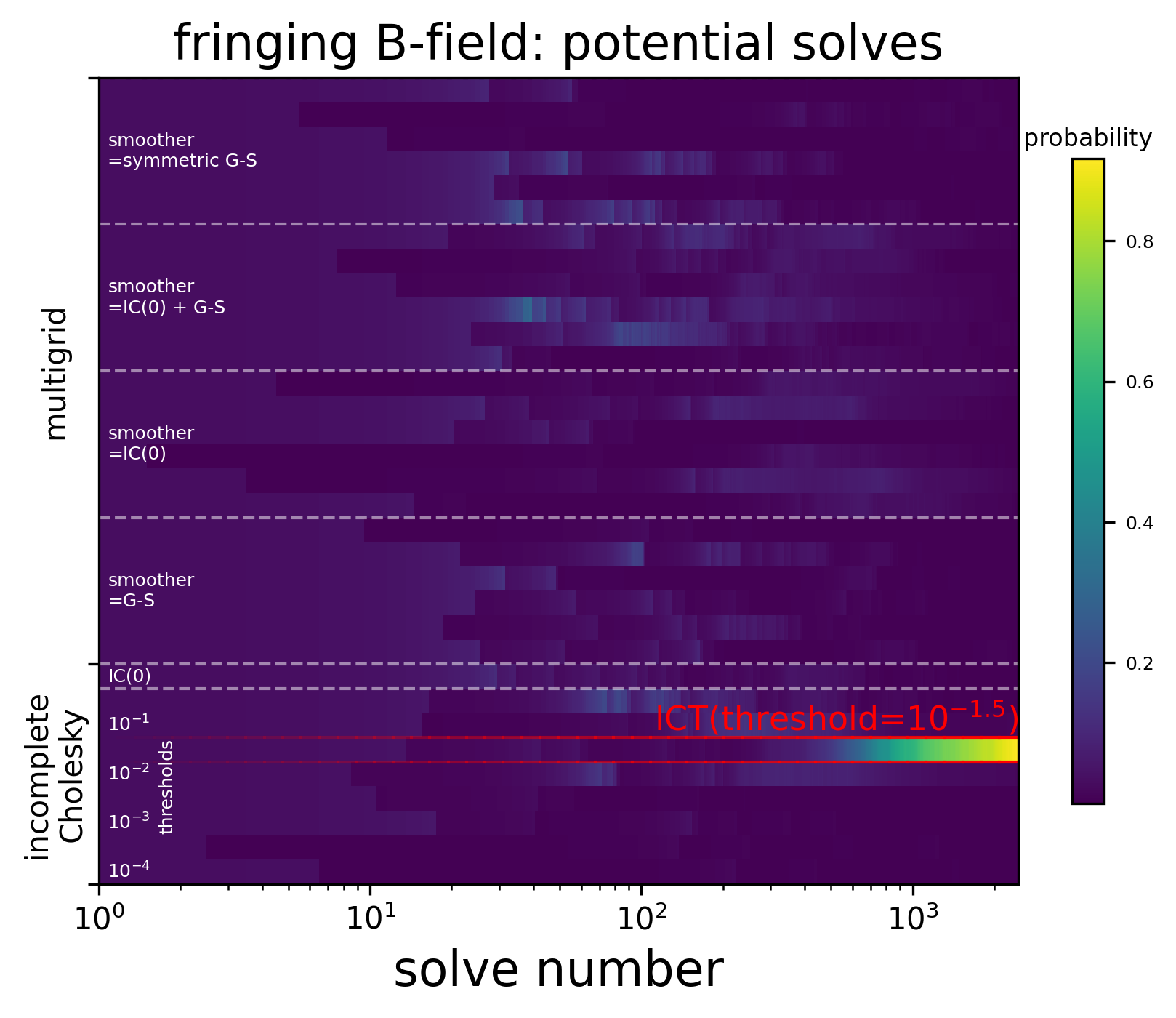}\ifdefined\arxiv\vspace{-3mm}\fi
	\caption{
		Heatmaps of probabilities assigned to each configuration at each solve of the \revision{pressure} equation~(left) and the electric potential~(right) on the fringing B-field simulation.
		The highlighted configuration is the one with the largest total probability across solves.\looseness-1
	}
	\label{fig:fringing-prob}
\end{minipage}

\end{figure}

In the case of the fringing B-field simulation, where PCGBandit performs much better, Figure~\ref{fig:fringing-prob} demonstrates more rapid convergence to a probability distribution dominated by a single configuration than in the case of Shercliff flow~(cf. Figure~\ref{fig:shercliff-prob}).
Furthermore, while Figure~\ref{fig:fringing-perf} shows no overall improvement over IC(0) on electric potential, in this simulation the compute is dominated by the \revision{pressure} solve~(cf. Table~\ref{tab:simulations}), which is significantly accelerated.\looseness-1

\subsection{Reproducible learning via deterministic cost estimates}\label{sec:deterministic}

\begin{figure}[!t]
	\begin{minipage}{0.5\linewidth}
		\includegraphics[width=\linewidth]{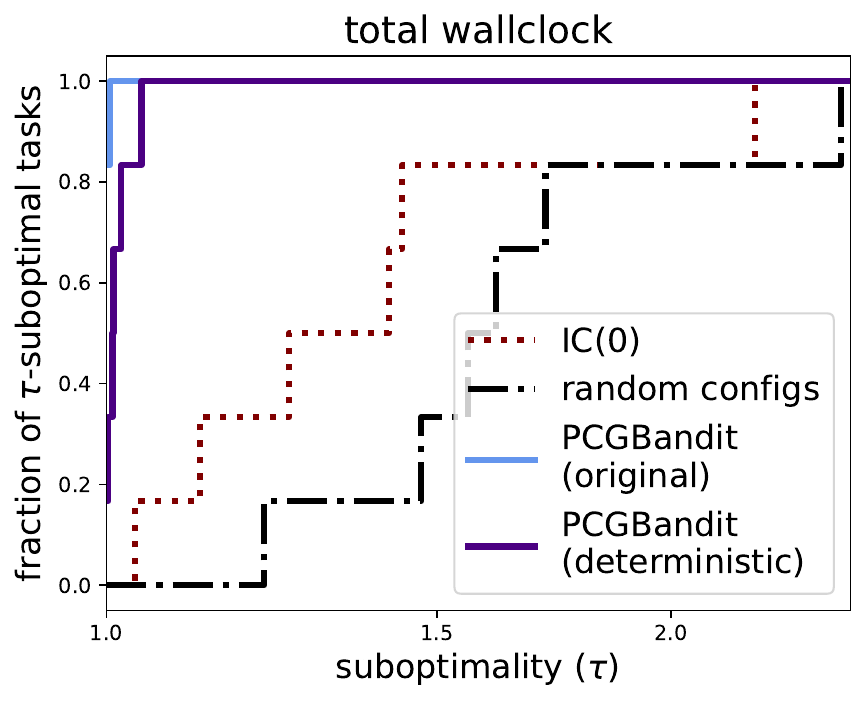}\ifdefined\arxiv\vspace{-3mm}\fi
		\caption{
			Performance profiles~\citep{dolan2002profiles} comparing regular PCGBandit and deterministic PCGBandit, where the latter uses Equation~\ref{eq:deterministic} to estimate the cost of each solve.
			We evaluate this in the setting of choosing between ICT thresholds, and so the baselines are IC(0) and sampling random configurations.
			The results demonstrate that using deterministic cost estimators is nearly as good as using the wallclock time.\looseness-1
		}
		\label{fig:perfprof-ICTC}
	\end{minipage}
	\hfill
	\begin{minipage}{0.45\linewidth}
		\includegraphics[width=\linewidth]{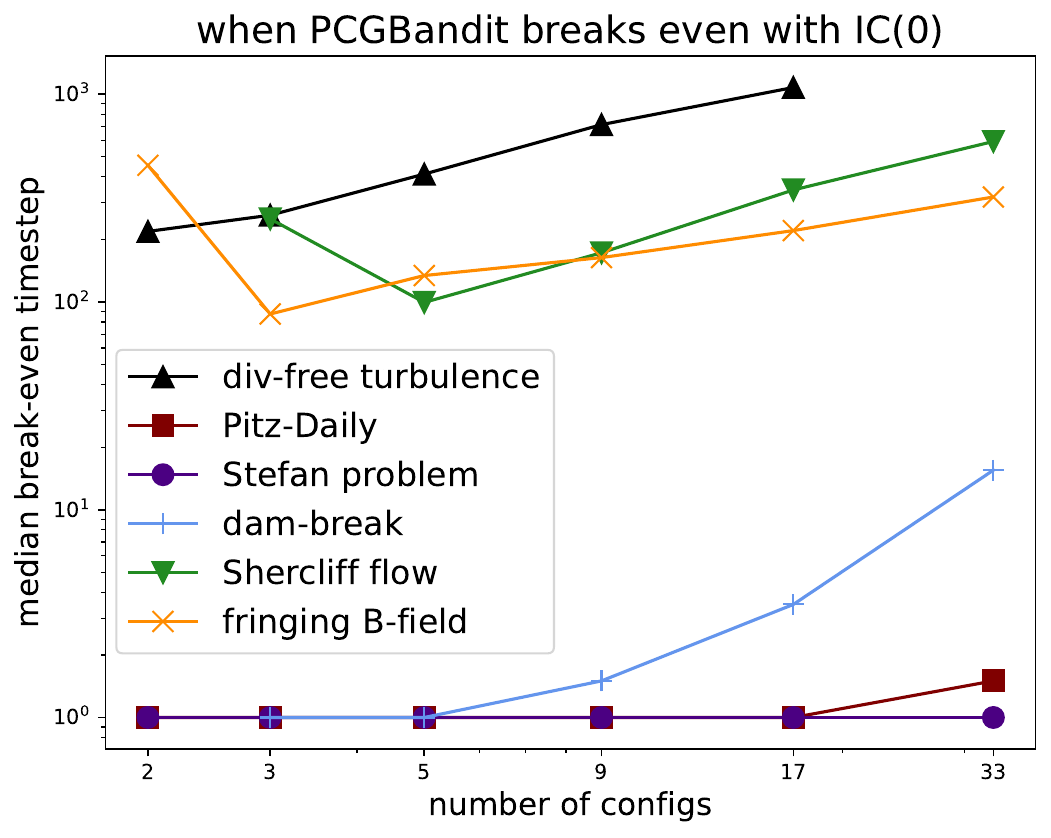}\ifdefined\arxiv\vspace{2.2mm}\fi
		\vspace{-7.5pt}
		\caption{The break-even timestep for each simulation, defined here as the timestep after which the {\em median across random seeds} of the cumulative cost of PCGBandit is smaller than that of the IC(0) baseline.
			As predicted by the $\BigO(\sqrt{dT})$ regret, the number of timesteps required to break even typically increases with the number of configurations.\looseness-1
		}
		\label{fig:breakeven-ICTC}
	\end{minipage}
\end{figure}

As noted in Section~\ref{sec:considerations}, using the wallclock cost of each solve as the bandit feedback leads to runs that are unreproducible even when fixing the random seed, which makes debugging difficult.
However, we can sometimes get a strong-enough signal from the number of PCG iterations taken by doing a rough analysis of the floating point operations~(FLOPs) used by the selected preconditioner at each step.
To test the efficacy of this, we run PCGBandit over a restricted search space of eight configurations---ICT with one of eight different thresholds or IC(0)\footnote{\revision{Our attempt at a FLOP-based multigrid performance estimator did not correlate well-enough with wallclock time; whether a good one is possible is left to future work.\looseness-1}}---and use the following estimate for the cost at time $t$ given the number of iterations $k_t$ that were required to solve $\*A_t\*x=\*b_t$:\looseness-1
\begin{equation}\label{eq:deterministic}
	(\nnz(\*A_t)+2\nnz(\*L_t)+5n)(10+k_t)
\end{equation}
Here $\nnz(\*A_t)$ is the number of nonzeros in the linear system and $\nnz(\*L_t)$ in the (threshold-dependent) lower triangular preconditioner, and  we add ten to the number of iterations to account for preconditioner construction costs.
As shown in Figure~\ref{fig:perfprof-ICTC}, this deterministic approach to cost estimation is competitive with using the wallclock time:
it does at most 1.2$\times$ worse on every task, which is significantly better than any baseline.

\subsection{The effect of the number of configurations}\label{sec:configs}

\begin{figure}[!t]
	\centering
	\includegraphics[width=0.325\linewidth]{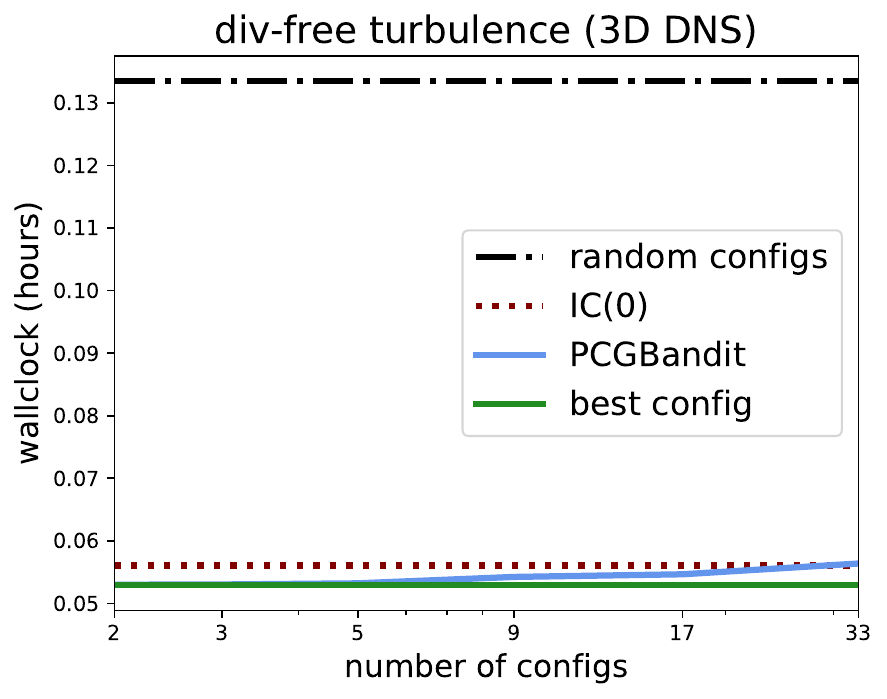}
	\includegraphics[width=0.325\linewidth]{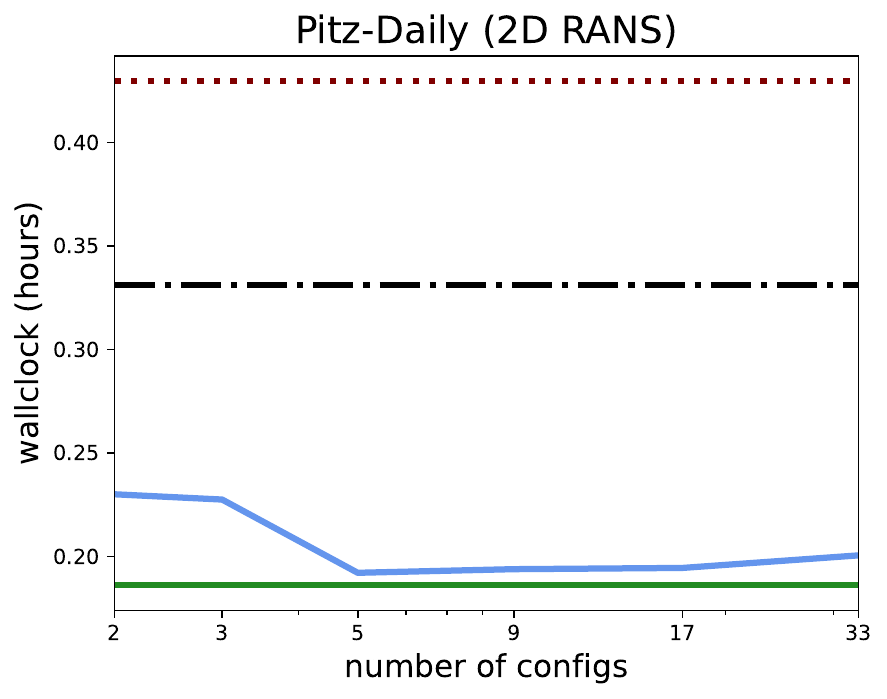}
	\includegraphics[width=0.32\linewidth]{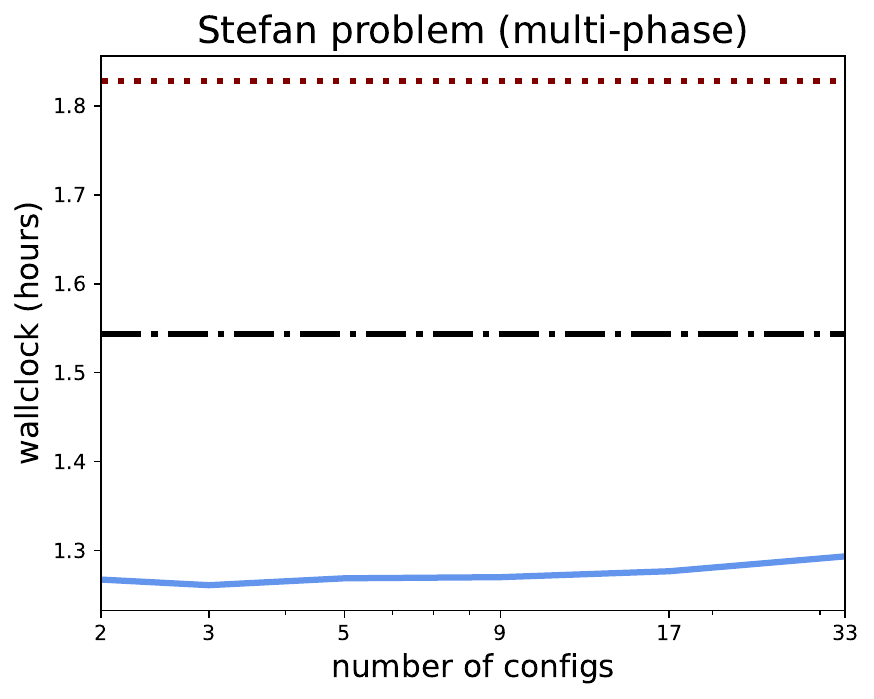}
	\includegraphics[width=0.325\linewidth]{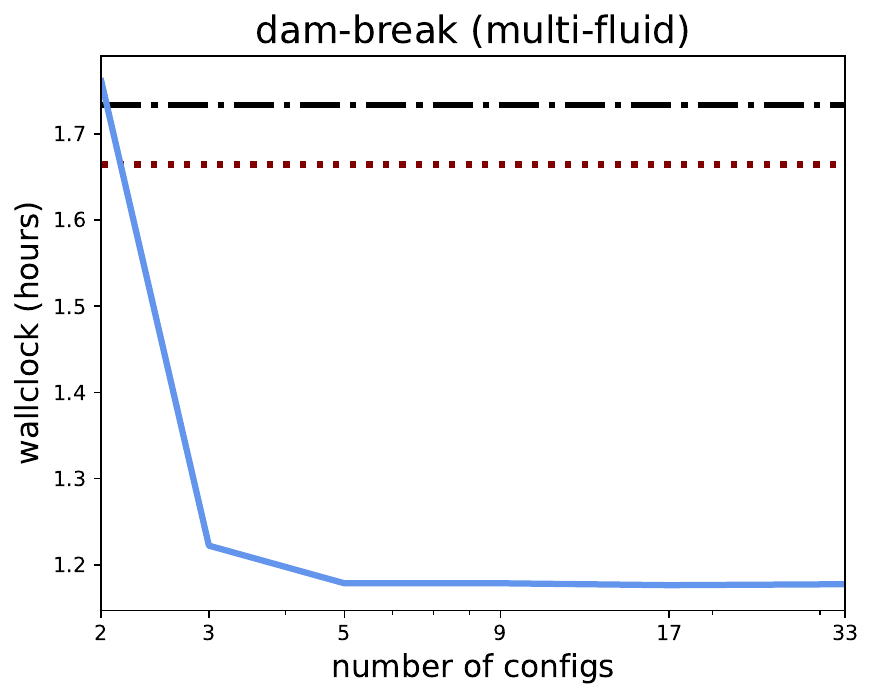}
	\includegraphics[width=0.325\linewidth]{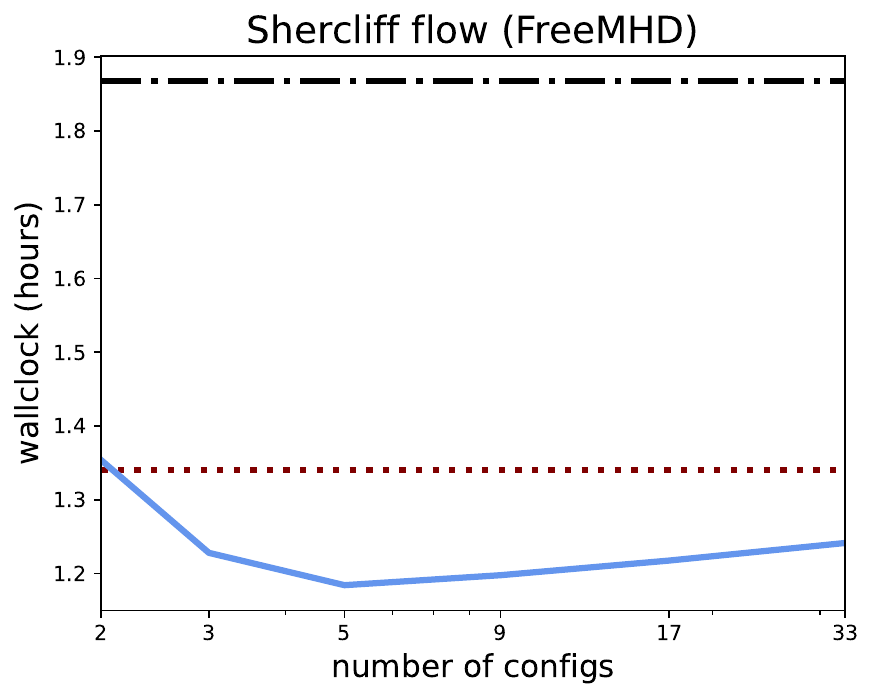}
	\includegraphics[width=0.32\linewidth]{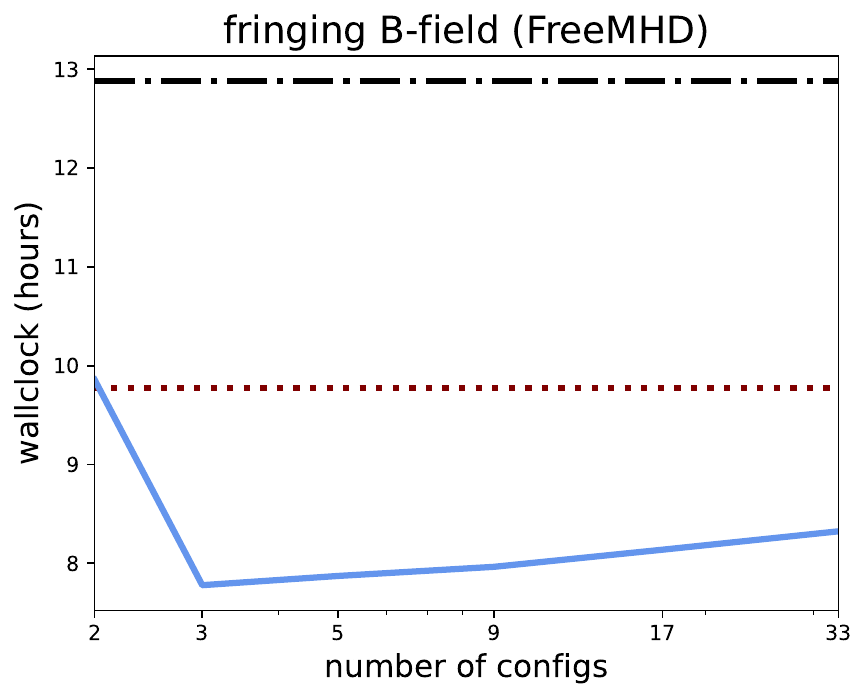}\ifdefined\arxiv\vspace{-3mm}\fi
	\caption{
		We consider using PCGBandit to tune the ICT threshold, with the configuration space thus containing only IC(0) and ICT with one to thirty-two different thresholds.
		Here we plot the performance of PCGBandit as a function of the number of thresholds~(i.e. configurations) considered.
	}
	\label{fig:ICTC}
\end{figure}

As discussed in Section~\ref{sec:algorithm}, the number of configurations $d$ considered by PCGBandit has a significant effect on the worst-case regret.
We now investigate how this manifests in practical simulations by using PCGBandit to tune only the ICT threshold while varying the number of thresholds under consideration~(we also keep the usual incomplete Cholesky preconditioner IC(0) as an option).
Specifically, we study the effect of learning across $2^p$ different thresholds for integers $p\in\{0,\dots,5\}$.
Figure~\ref{fig:ICTC} demonstrates that the number of thresholds considered has in some cases a dramatic effect on the performance, but the best performance is typically obtained at five or nine configurations~(i.e. four or eight thresholds plus IC(0)).
As we might expect from the worst-case analysis, Figure~\ref{fig:breakeven-ICTC} also shows that on the harder~(for PCGBandit) simulations the break-even point is typically increasing as a function of the number of configurations $d$.

\section{Conclusion}

In this paper we demonstrate how machine learning can consistently accelerate existing, widely used numerical simulation software, with no hidden upfront or overhead costs due to pretraining or model size.
We do this by integrating standard online learning techniques into OpenFOAM's linear system solver by developing PCGBandit, which tunes the solver's preconditioners using its own wallclock as feedback.
In certain settings, this leads to substantial wallclock reductions in the overall simulation wallclock.

There are many directions for further development of learning-enhanced numerical simulation.
This includes algorithmic work on better bandit algorithms and making use of context, extending the results to other linear solvers such as GMRES or even nonlinear ones, designing better configuration spaces, and making use of multiple instances to find better initializations.
Further afield, methods that make better use of the specific linear solvers can be developed, e.g. to initialize Krylov subspaces.

\section*{Acknowledgments}

\ifdefined\anon
Removed for anonymity.
\else
This research is supported by Korea Institute of Fusion Energy~(KFE) R\&D Program CN2502-1, MSIT National Research Foundation of Korea~(NRF) Grant No. RS-2024-00358933, and US Department of Energy (DOE) Grant No. DESC0024626. Mikhail Khodak acknowledges the resources of the Princeton Language and Intelligence unit of the Princeton AI Lab. Min Ki Jung acknowledges the Research Institute of Energy and Resources and the Institute of Engineering Research at Seoul National University.
\fi

\bibliographystyle{elsarticle-harv}
\bibliography{refs}

\newpage
\appendix

\section{Additional figures}

\begin{figure}[!h]
	\centering
	\includegraphics[width=0.495\linewidth]{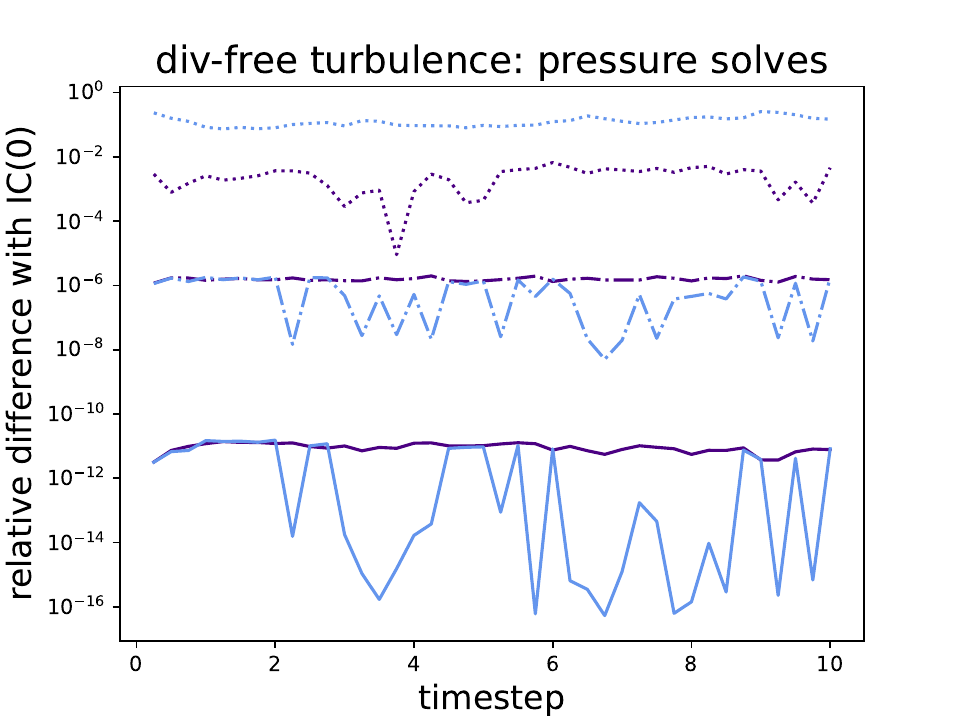}
	\includegraphics[width=0.495\linewidth]{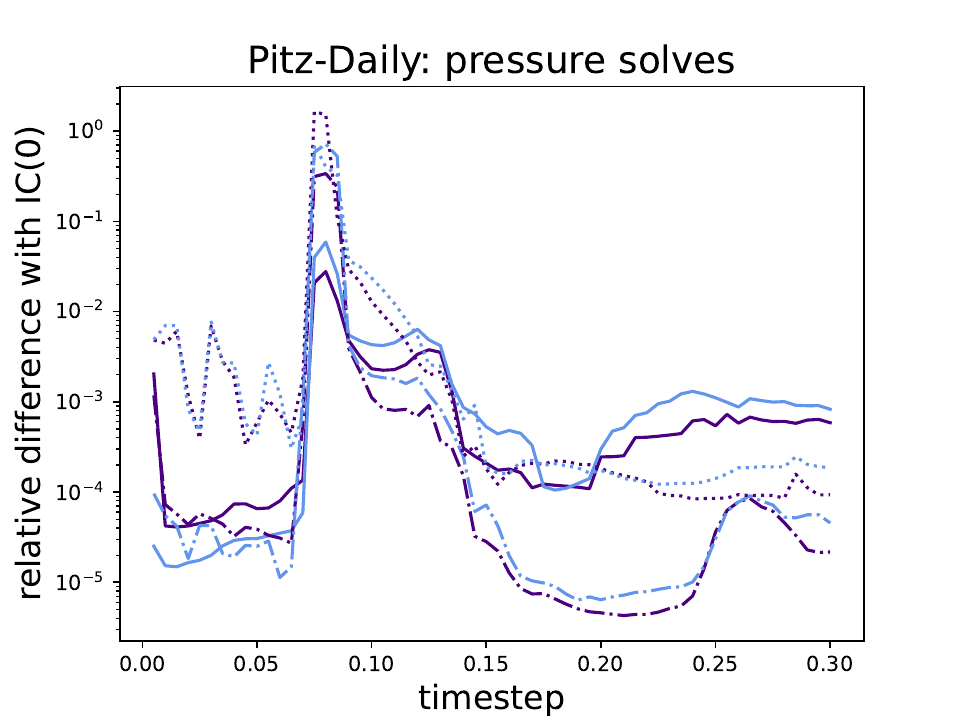}
	\includegraphics[width=0.495\linewidth]{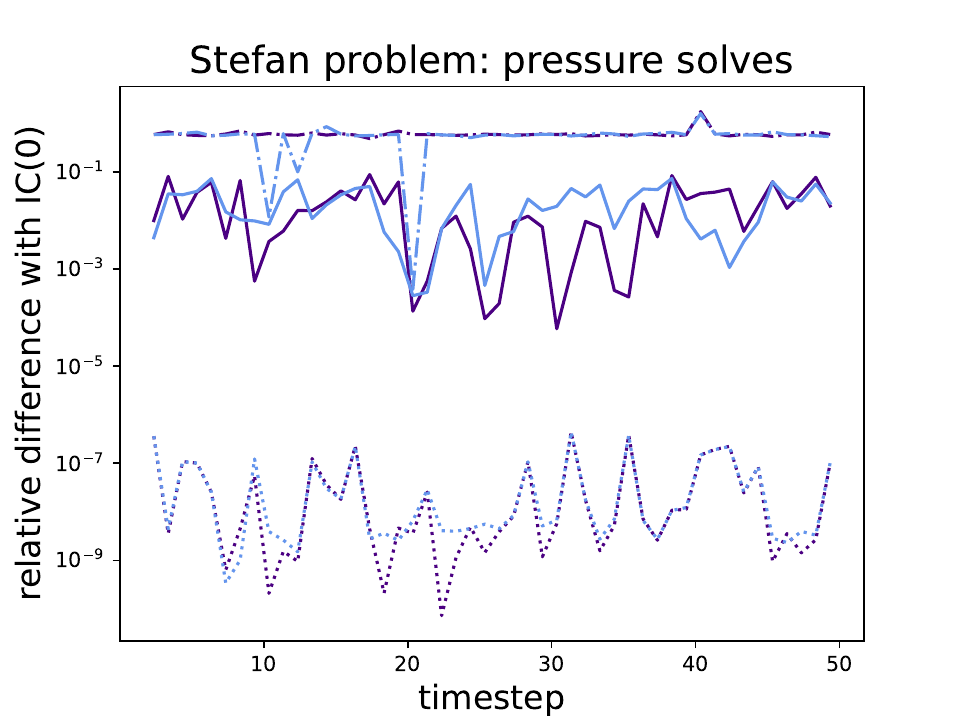}
	\includegraphics[width=0.495\linewidth]{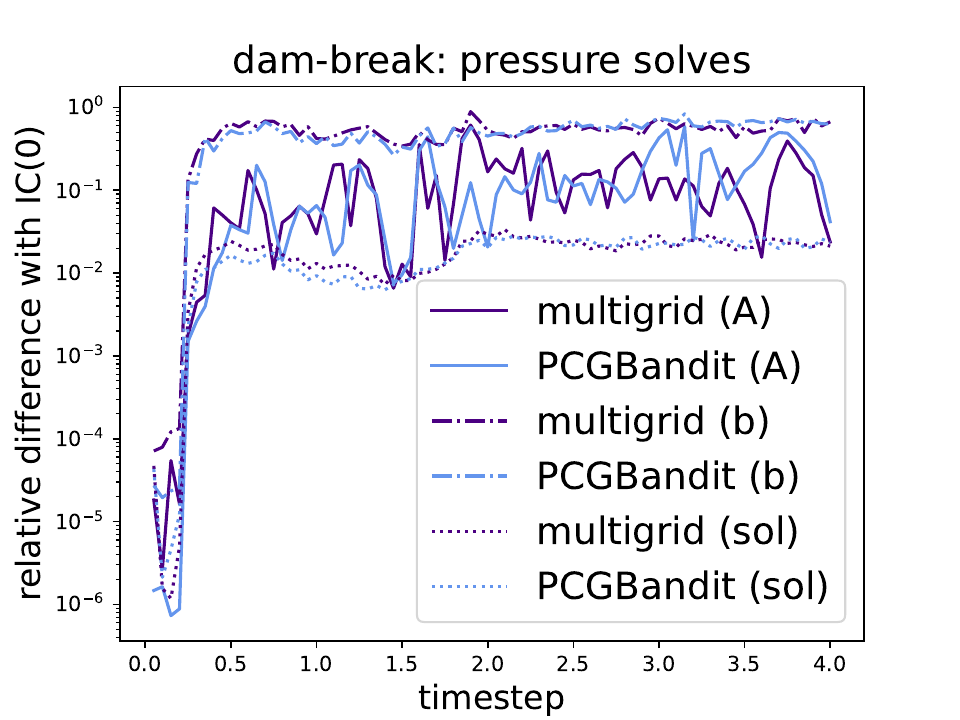}
	\includegraphics[width=0.495\linewidth]{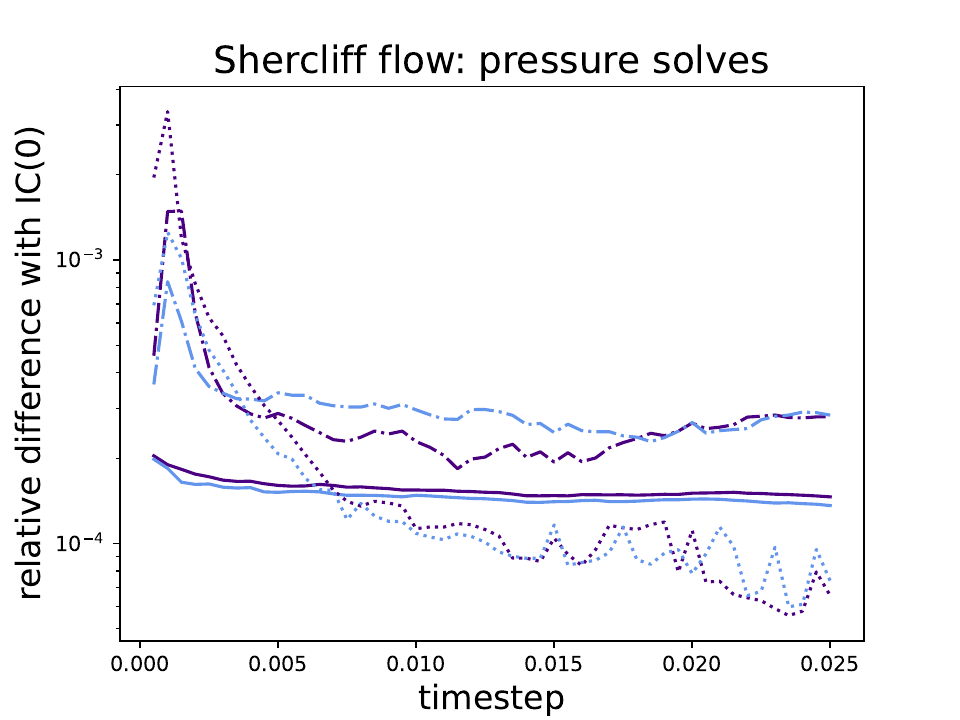}
	\includegraphics[width=0.495\linewidth]{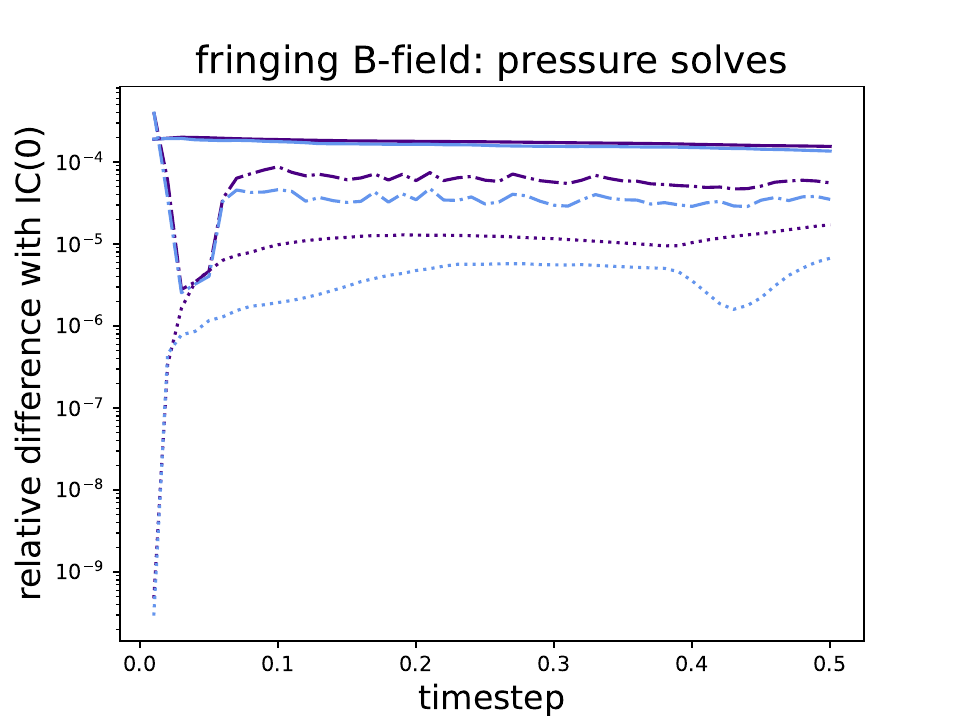}\ifdefined\arxiv\vspace{-3mm}\fi
	\caption{
		\revision{Relative distances from the linear systems (and their solutions) obtained by always using IC(0) preconditioning to those obtained by either always using multigrid or selecting preconditioners using PCGBandit.
		While distances between solutions are typically small, due to ill-conditioning they can be nonnegligible in some cases (e.g. of divergence-free turbulence), even when the linear systems are close.}
		\looseness-1
	}
	\label{fig:diff}
\end{figure}

\end{document}